\documentclass[sensors,article,accept,moreauthors,pdftex]{Definitions/mdpi}
\pdfoutput=1

\firstpage{1}
  \articlenumber{2948}
\pubvolume{22}
\issuenum{8}
\pubyear{2022}
\copyrightyear{2022}
\externaleditor{Academic Editors: Sung-Phil~Kim and Yvonne~Tran}
\datereceived{6 February 2022}
\dateaccepted{10 April 2022}
\datepublished{12 April 2022}
\hreflink{https://doi.org/10.3390/s22082948}
\usepackage[T5,T1]{fontenc}
\usepackage[russian,english]{babel}
\usepackage{pifont}
\usepackage{relsize}

\Title{Automatic Muscle Artifacts Identification and Removal \mbox{from Single-Channel} EEG Using Wavelet Transform with Meta-Heuristically Optimized Non-Local Means Filter}

\TitleCitation{Automatic Muscle Artifacts Identification and Removal from Single-Channel EEG Using Wavelet Transform with Meta-Heuristically Optimized Non-Local Means Filter}

\Author{Souvik~Phadikar~\textsuperscript{1}\textsuperscript{,}*\href{https://orcid.org/0000-0002-7122-2095}{\orcidicon}, Nidul~Sinha~\textsuperscript{1}, Rajdeep~Ghosh~\textsuperscript{2}\href{https://orcid.org/0000-0002-8045-0137}{\orcidicon} and Ebrahim~Ghaderpour~\textsuperscript{3}\href{https://orcid.org/0000-0002-5165-1773}{\orcidicon}}

\AuthorNames{Souvik Phadikar, Nidul Sinha, Rajdeep Ghosh and Ebrahim Ghaderpour}

\AuthorCitation{Phadikar, S.; Sinha, N.; Ghosh, R.; Ghaderpour, E.}

\address{\textsuperscript{1} \quad Department of Electrical Engineering, National Institute of Technology Silchar, Silchar 788010, Assam, India; nidulsinha@ee.nits.ac.in

\textsuperscript{2} \quad Department of Information Technology, Gauhati University, Guwahati 781014, Assam, India; rajdeepghosh@gauhati.ac.in

\textsuperscript{3} \quad Department of Geomatics Engineering, University of Calgary, Calgary, AB T2N 1N4, Canada; ebrahim.ghaderpour@ucalgary.ca}

\corres{Correspondence: souvik\_rs@ee.nits.ac.in}

\abstract{Electroencephalogram (EEG) signals may get easily contaminated by muscle artifacts, which may lead to wrong interpretation in the brain--computer interface (BCI) system as well as in various medical diagnoses. The main objective of this paper is to remove muscle artifacts without distorting the information contained in the EEG. A novel multi-stage EEG denoising method is proposed for the first time in which wavelet packet decomposition (WPD) is combined with a modified non-local means (NLM) algorithm. At first, the artifact EEG signal is identified through a pre-trained classifier. Next, the identified EEG signal is decomposed into wavelet coefficients and corrected through a modified NLM filter. Finally, the artifact-free EEG is reconstructed from corrected wavelet coefficients through inverse WPD. To optimize the filter parameters, two meta-heuristic algorithms are used in this paper for the first time. The proposed system is first validated on simulated EEG data and then tested on real EEG data. The proposed approach achieved average mutual information (MI) as 2.9684 $\pm$ 0.7045 on real EEG data. The result reveals that the proposed system outperforms recently developed denoising techniques with higher average MI, which indicates that the proposed approach is better in terms of quality of reconstruction and is fully automatic.}


\makeatletter
\DeclareRobustCommand*\textsubscript[1]{%
  \@textsubscript{\selectfont#1}}
\def\@textsubscript#1{%
  {\m@th\ensuremath{_{\mbox{\fontsize\sf@size\z@#1}}}}}
\makeatother

\usepackage{sansmath}

\begin{document}
\section{Introduction \label{sect:sec1-sensors-1607106}}

Brain--computer interface (BCI) is a collaborative setup between the human brain and a computer device, which takes brain signals as input and tries to decode them into computer commands to direct external activities such as cursor control, wheelchair control, and \mbox{silent-speech} recognition~\cite{B1-sensors-1607106,B2-sensors-1607106}. The electroencephalogram (EEG) is widely used as a \mbox{non-invasive} way of measuring brain signals in BCI systems due to its high temporal resolutions. It captures cerebral activities through electrodes placed over the scalp. EEG is also used in diagnosing several neurological disorders such as Alzheimer’s and epileptic seizures~\cite{B3-sensors-1607106}. Generally, BCI records the brain activities, pre-processes them to extract features from the data and finally classifies them for identifying mental states~\cite{B4-sensors-1607106}. Therefore, analyzing the information contained in the EEG is of great importance. Although EEG is intended to record cerebral activities in the form of electrical pulses, it also captures electrical activities other than the cerebral activities, known as artifacts. EEG signals are widely contaminated with different types of artifact sources such as cardiogenic (ECG), ocular (EOG), and myogenic (EMG). These artifacts suppress the information contained in the EEG signal and lead to wrong interpretation of BCI systems and medical diagnosis. Hence, the elimination of artifacts from the EEG signal is of great importance for \mbox{practical uses}.

The elimination of EMG artifacts from an EEG is more challenging as compared to removing other types of artifacts from an EEG signal. Generally, the EMG artifacts occur in EEG signals due to various muscle movements near the head. Various muscle movements such as teeth clenching, swallowing, head movement, chewing, jaw movement, and tongue movement are also captured through electrodes and contaminate the EEG signals with higher amplitude, broad anatomical distribution, and wide frequency spectrum~\cite{B5-sensors-1607106}. Specifically, the EMG has a wide spectral band and broad spatial distribution~\cite{B6-sensors-1607106}. The power-frequency spectrum of EMG artifacts ranges from 2 Hz to 100 Hz~\cite{B6-sensors-1607106}. Their wide spectrum easily overlaps with the frequencies of interest of the EEG. Because of the influence of both the volume conduction and the broad distribution of muscles across the face, neck, and head, it can be observed anywhere on the scalp.

In contrast, the spatial distribution of ECG and EOG are comparatively localized. EMG artifacts occur from the movement of spatially distributed and functionally independent muscle groups with distinct topographic and spectral signatures~\cite{B7-sensors-1607106}. The spectral signatures may vary across different muscles and also with the intensity of muscle contraction. In addition, ECG and EOG artifacts can be removed with the help of a reference channel, while it is difficult to remove EMG artifacts by using reference channels. Due to the complex muscle distribution, the placement of reference electrode for EMG artifact removal is very difficult~\cite{B7-sensors-1607106} in a practical situation. Hence, it is challenging to remove EMG artifacts without using any reference channels.

\textls[-10]{Numerous techniques have been proposed in the literature to overcome these difficulties of EMG removal from EEG. In early research, approaches based on blind source separation (BSS) were explored for the automatic removal of EMG artifacts from \mbox{multichannel EEG~\cite{B8-sensors-1607106}.} The authors in~\cite{B9-sensors-1607106,B10-sensors-1607106,B11-sensors-1607106} showed that independent component analysis (ICA) resulted in a good performance for removing EMG artifacts from the multichannel EEG data. \mbox{Janani et~al.~\cite{B12-sensors-1607106}} showed that canonical correlation analysis (CCA) outperformed the ICA to eliminate EMG from EEG successfully. Chen et~al.~\cite{B13-sensors-1607106} proved that independent vector analysis (IVA) could successfully denoise the EEG signal from EMG. However, BSS-based denoising techniques assume that the number of concerned sources is equal to or less than the number of channels. In this case, BSS can also be implemented for single-channel EEG data~\cite{B14-sensors-1607106}, where \mbox{single-channel} EEG data was at first decomposed with some decomposition techniques. Then artifactual components from the results of decomposition are used for BSS processing.}

However, the recent trend in mobile healthcare applications has led to the reduction in the number of EEG channels for health monitoring, and in some cases only a single channel is used~\cite{B15-sensors-1607106}. In such applications, BSS may not perform well. To overcome this issue, several attempts based on the decomposition techniques have been proposed to remove EMG artifacts from a few channels and single-channel EEG data. \mbox{Fitzgibbon et~al.~\cite{B16-sensors-1607106}} removed EMG artifacts from central channels through the surface Laplacian transform (SLT). They stated that brain activities captured through central channels are affected only by a group of nonadjacent muscles as there is no muscle under the center of the scalp~\cite{B16-sensors-1607106} and SLT performed well in removing muscle contamination of scalp signals. However, the neuronal activities from other channels are affected by both the adjacent and the distant group of muscles. Yong et~al.~\cite{B17-sensors-1607106} removed EMG artifacts from both the non-central and distant channels through morphological component analysis (MCA) by assuming that the EEG recordings are a linear combination of neuronal activities, artifacts, and electronic noise. However, as the EEG and EMG have non-stationary morphology, the performance of the MCA is not always satisfactory with the chosen dictionaries in MCA. Further, the above-mentioned decomposition techniques are not fully data-driven, which make them unfit for automatic and online applications. Bhardwaj et~al.~\cite{B18-sensors-1607106} proposed wavelet-transform-based EMG artifact removal. Multiscale principal component analysis (MSPCA) is another effective hybrid denoising technique for EMG artifacts correction, in which PCA is combined with multiscale wavelet transform~\cite{B19-sensors-1607106,B20-sensors-1607106,B21-sensors-1607106}. Wavelets separate stochastic and deterministic processes and approximately uncorrelated the autocorrelation between calculations, whereas PCA offers interactions between distinct variables. The wavelet transform is an ideal approach for EEG signal analysis due to its \mbox{multi-resolution characteristics.}

In recent years, several hybrid methods~\cite{B12-sensors-1607106,B22-sensors-1607106,B23-sensors-1607106,B24-sensors-1607106,B25-sensors-1607106,B26-sensors-1607106,B27-sensors-1607106,B28-sensors-1607106,B29-sensors-1607106,B30-sensors-1607106,B31-sensors-1607106,B32-sensors-1607106} have been proposed to overcome the difficulties of individual techniques for the removal of EMG artifacts from the EEG data, and their summary is shown in \ref{tabref:sensors-1607106-t001}. It is observed from the table that \mbox{EMD-based} methods have been widely used for EMG artifacts correction. However, these methods ignore significant cross-channel interdependence information and consider the dependent information between spatially adjacent channels for the source separation procedure~\cite{B33-sensors-1607106}. On the other hand, EMD-based techniques perform well on few channel EEG data compared to the single-channel EEG data, whereas wavelet-based techniques perform well on single-channel EEG data. The main idea of wavelet-based techniques is to threshold the wavelet coefficients to remove the artifacts from the signal. To threshold the wavelet coefficients, several thresholding techniques such as statistical threshold and universal threshold calculations are widely used. However, Phadikar et~al.~\cite{B34-sensors-1607106} showed that these threshold functions may vary with the data, which makes the system unfit for automatic and online implementation. Recently, Bajaj et~al.~\cite{B35-sensors-1607106} proposed a tuneable artifact removal technique based on wavelet packet decomposition (WPD). Their denoising method automatically removes EMG artifacts from the corrupted EEG signal. However, the challenge associated with wavelet analysis remains in the selection of proper wavelet function and decomposition level.  
\vspace{-3pt}  
    \begin{table}[H]
    \tablesize{\small}
    \caption{Comparison of recently developed hybrid techniques for the removal of EMG artifacts from the EEG data.}
 
\begin{adjustwidth}{-\extralength}{0cm}
   \label{tabref:sensors-1607106-t001}

\setlength{\cellWidtha}{\fulllength/7-2\tabcolsep+0.3in}
\setlength{\cellWidthb}{\fulllength/7-2\tabcolsep+0.3in}
\setlength{\cellWidthc}{\fulllength/7-2\tabcolsep-0in}
\setlength{\cellWidthd}{\fulllength/7-2\tabcolsep-0.2in}
\setlength{\cellWidthe}{\fulllength/7-2\tabcolsep-0.2in}
\setlength{\cellWidthf}{\fulllength/7-2\tabcolsep-0.2in}
\setlength{\cellWidthg}{\fulllength/7-2\tabcolsep-0in}
\scalebox{1}[1]{\begin{tabularx}{\fulllength}{>{\centering\arraybackslash}m{\cellWidtha}>{\centering\arraybackslash}m{\cellWidthb}>{\centering\arraybackslash}m{\cellWidthc}>{\centering\arraybackslash}m{\cellWidthd}>{\centering\arraybackslash}m{\cellWidthe}>{\centering\arraybackslash}m{\cellWidthf}>{\centering\arraybackslash}m{\cellWidthg}}
\toprule

\textbf{Author(s)} & \textbf{Methodology} & \textbf{Requirement of Reference Channel} & \textbf{Prior Knowledge} & \textbf{Automatic} & \textbf{Online} & \textbf{Can Perform on Single Channel}\\
\cmidrule{1-7}

Mijovic et~al. {\cite{B22-sensors-1607106}} & EEMD-ICA & \ding{55} & \ding{55} & \ding{55} & \ding{55} & \ding{51}\\

Sweeney et~al. {\cite{B23-sensors-1607106}} & EEMD-CCA & \ding{55} & \ding{55} & \ding{55} & \ding{55} & \ding{51}\\

Chen et~al. {\cite{B24-sensors-1607106}} & EEMD-IVA & \ding{55} & \ding{55} & \ding{55} & \ding{55} & \ding{51}\\

Chen et~al. {\cite{B25-sensors-1607106}} & EEMD-MCCA & \ding{55} & \ding{55} & \ding{55} & \ding{55} & \ding{51}\\

Chen et~al. {\cite{B14-sensors-1607106}} & MEEMD-CCA & \ding{55} & \ding{55} & \ding{55} & \ding{55} & \ding{51}\\

Zeng et~al. {\cite{B26-sensors-1607106}} & MEEMD-ICA & \ding{55} & \ding{55} & \ding{55} & \ding{55} & \ding{55}\\

Chen et~al. {\cite{B27-sensors-1607106}} & MEMD-CCA & \ding{55} & \ding{55} & \ding{55} & \ding{55} & \ding{55}\\

Xu et~al. {\cite{B28-sensors-1607106}} & MEMD-IVA & \ding{55} & \ding{55} & \ding{55} & \ding{55} & \ding{55}\\

Maddirala et~al. {\cite{B29-sensors-1607106}} & SSA-ICA & \ding{55} & \ding{55} & \ding{55} & \ding{55} & \ding{51}\\

Dora et~al. {\cite{B30-sensors-1607106}} & Adaptive SSA-NNR & \ding{51} & \ding{51} & \ding{51} & \ding{55} & \ding{51}\\

Liu et~al. {\cite{B31-sensors-1607106}} & FMEMD-CCA & \ding{55} & \ding{55} & \ding{55} & \ding{55} & \ding{55}\\

Li et~al. {\cite{B32-sensors-1607106}} & cICA & \ding{51} & \ding{51} & \ding{51} & \ding{55} & \ding{55}\\

\bottomrule
\end{tabularx}}
\end{adjustwidth}

    \noindent\footnotesize{\ding{55}---No, \ding{51}---Yes; EEMD---Ensemble Empirical Mode Decomposition, MCCA---Multiset Canonical Correlation Analysis, MEEMD---Multivariate EEMD, SSA---Singular Spectrum Analysis, NNR---Neural Network Regressor, FMEMD---Fast Multivariate Empirical Mode Decomposition, cICA---Constrained ICA. }
    \end{table}
    \vspace{-6pt}

Zhang et~al.~\cite{B36-sensors-1607106} proposed a convolutional neural network-based approach for the removal of muscle artifacts from the EEG signals. However, their approach is limited to only 2 s of EEG epoch. A better configuration of CNN is needed for longer duration of EEG. More recently, NLM filter-based methods were employed for the removal of EMG artifacts from the EEG data. Earlier it was designed for image denoising~\cite{B37-sensors-1607106}. \mbox{Eltrass et~al.~\cite{B38-sensors-1607106}} proposed a hybrid method in which an NLM filter is combined with the multi-kernel normalized least mean square with a coherence-based sparsification (MKNLMS-CS) algorithm for the successful removal of the EMG artifacts from the EEG data.

While numerous methods have been studied for the elimination of EMG, developing robust single or hybrid algorithms that can operate automatically is still very challenging. In addition, most of the techniques are suitable for multi-channel EEG (performance degrades for single-channel EEG), hence, developing an automatic EMG artifact removal technique for single-channel EEG is still challenging. However, the combination of multiple cascading algorithms is essentially an unexplored area. The significance of cascading algorithms is to suppress the artifacts in a single stage, which suppresses artifact sources and achieves a higher degree of robustness. In this manuscript, a novel automatic cascaded system is proposed in which wavelet transform is combined with the NLM filter for the elimination of EMG artifacts from the EEG signals.

In the proposed work, an EMG-contaminated EEG signal is automatically identified and decomposed into wavelet coefficients through WPD. Wavelet transform is used to separate the EEG features into different scales so that significant features of the EEG signals are preserved and noises can be removed. The mother wavelet and with an appropriate decomposition level is selected through a proper procedure. It is to be noted that artifacts will be reflected in the wavelet coefficients. Hence, correcting the corrupted wavelet coefficients will result in denoising the EEG signal. In the proposed method, corrupted wavelet coefficients are corrected through an optimized NLM algorithm. Finally, all the corrected coefficients are used in the inverse operation to get back the original artifact-free EEG signal. In this approach, no threshold calculation method is employed; instead, wavelet coefficients are corrected through an NLM filter, which makes the system automatic and fit for implementation in online applications. In addition, a modified NLM algorithm is proposed, in which all the parameters of the algorithm are optimized in a proper way. The proposed approach preserves the information of interest in the EEG signal, which is reflected in higher average correlation coefficients (CC) and higher MI values. The main contributions of the proposed denoising algorithm are stated as:\begin{enumerate}
\item[(1)] The proposed algorithm employs the wavelet transform for its good time-frequency localization and optimized NLM filter for better removal of wide frequency spectrum EMG artifacts from the corrupted EEG signals.
\item[(2)] The main challenge in wavelet-transform-based denoising is the selection of appropriate threshold values. Hence, instead of thresholding the wavelet coefficients, they are corrected through NLM estimation.
\item[(3)] The issue of optimum parameter selection for the NLM is properly addressed with the help of a meta-heuristic algorithm.
\item[(4)] Further, the proposed algorithm corrects only the EMG-corrupted portion of the EEG and keeps the clean portion untouched. Hence, the proposed system preserves the true information contained in the EEG signal.
\end{enumerate}

The novelty of the proposed hybrid method in which SVM, WPD and NLM are combined is described as below:\begin{enumerate}[label=$\bullet$]
\item SVM is one of the most efficient classifiers and hence was used in this work. Different artifacts have different characteristics, for example, eyeblink artifacts contaminate the EEG signals with 5 to 10 times’ larger amplitude than the normal EEG signals with very low frequency (typically 0.1 to 3 Hz), and EMG artifacts contaminate the EEG signals with higher amplitude and higher frequency components. Similarly, ECG artifacts contaminate the EEG signals with periodic discharges. Moreover, we do not know which EEG patterns are hidden in the artifact portions (there is no ground truth). Hence, before removal of artifacts, it is necessary to identify which artifacts contaminate the EEG signals, because different artifacts have different characteristics and affect the EEG signals distinctively. If the system does not have any prior knowledge about the type of artifacts, then the system will be capable of denoising the EEG signals but the true EEG pattern (or the original ground truth) will not be reconstructed properly. Therefore, first the EEG signals corrupted with EMG artifacts are identified with SVM as a classifier.
\item The frequency of muscle artifacts overlaps with the frequency of interest of EEG signals. Hence, removing those particular frequencies by making the relevant wavelet coefficients zero will result in a significant loss of information. Therefore, at first EEG signals are decomposed into wavelet coefficients using WPD to represent the different frequency bands of EEG signals with respect to time. Note that designing an efficient NLM-based filter for a very wide frequency band is very difficult. Thus, the corrupted EEG signal is decomposed into different frequency bands at different levels in WPD so that efficient NLM filters can be designed for decomposed signals at different levels for the effective removal (correction) of EMG artifacts.
\item The optimization of parameters, especially the bandwidth parameter, $\uplambda$ of NLM algorithm at different levels (set of $\uplambda$), simultaneously is a challenging task, and meta-heuristic algorithms are the most efficient ones in solving this type of problem. Hence, one of the recently developed meta-heuristic algorithms, GWO, is used for the task.
\end{enumerate}

The manuscript is organized as: Section \ref{sect:sec1-sensors-1607106} introduces the background, state-of-arts, challenges, objectives, and notations and preliminaries used in this paper are described in \ref{tabref:sensors-1607106-t002}, Section \ref{sect:sec2-sensors-1607106} depicts the tools and ideas employed in this methodology, Section \ref{sect:sec3-sensors-1607106} states the proposed methodology, Section \ref{sect:sec4-sensors-1607106} depicts the outcome of the research, Section \ref{sect:sec5-sensors-1607106} discusses about the experiment, and Section \ref{sect:sec6-sensors-1607106} completes the manuscript by drawing conclusions alongside the strategies for the future works.    
\vspace{-3pt}
    \begin{table}[H]
    \tablesize{\small}
    \caption{Notations and preliminaries.}
    \label{tabref:sensors-1607106-t002}

\setlength{\cellWidtha}{\textwidth/2-2\tabcolsep-0.5in}
\setlength{\cellWidthb}{\textwidth/2-2\tabcolsep+0.5in}
\scalebox{1}[1]{\begin{tabularx}{\textwidth}{>{\raggedright\arraybackslash}m{\cellWidtha}>{\raggedright\arraybackslash}m{\cellWidthb}}
\toprule

\textbf{Name} & \textbf{Details}\\
\cmidrule{1-2}

cA\textsubscript{j}(k) & Approximation Coefficients at level j and instant k.\\

cD\textsubscript{j}(k) & Detail Coefficients at level j and instant k.\\

x(t) or \emph{X} & Input EEG signal\\

X\textsuperscript{\^{}} & Reconstructed clean EEG signal\\

Y & Simulated Corrupted EEG\\

$y^{*} $ & Simulated clean EEG\\

        h(k)
       & Highpass Filter in Wavelet Packet Decomposition\\

        g(k)
       & Lowpass Filter in Wavelet Packet Decomposition\\

$\textit{d}_{\textit{i},\textit{j}} $ & Wavelet Coefficient at \emph{i}th level and \emph{j}th node.\\

$\textit{d}_{\textit{i},\textit{j}}^{\hat{}} $ & Corrected wavelet coefficients\\

$\upvarepsilon $ & Reconstruction Error\\

$\textit{X}_{\textit{r}\textit{e}\textit{c}\textit{o}\textit{n}} $ & Reconstructed signal\\

$\textit{S}\textit{E}_{\textit{j}} $ & Normalized Shannon Entropy at level \emph{j}\\

$\textit{E}_{\textit{j}\textit{k}} $ & Wavelet Energy Spectrum at level \emph{j} and instant \emph{k}\\

$\widetilde{\textit{u}}\left( \textit{s} \right) $ & Estimated version of the signal \emph{u}\\

\emph{N}(\emph{S})
       & Search Neighbourhood\\

$W$($S,~\eta$)
       & Weights corresponding to given sample \emph{S}\\

\emph{$\lambda$} & Bandwidth Parameter\\

        P
       & Patch half-width\\

       M
       & Search neighbourhood half-width\\

        $\text{$\Delta$}$
       & Patch\\

$\textit{L}_{\Delta} $ & Number of samples contained in $\text{$\Delta$}$\\

$\textit{d}^{2}\left( {\textit{s},\textit{t}} \right) $ & Squared-summation of the point-by-point difference between patches\\

        C-EEG
       & Corrupted EEG signal\\

        NC-EEG
       & Non-corrupted EEG signal\\

       P (,)
       & Joint Probability Distribution Function\\

        P ( )
       & Marginal Probability Distribution Function\\

       SAR
       & Signal to Artifact Ratio\\

$\textit{W}^{- 1} $ & Inverse WPD Function\\

$\mu $ & Sample Mean\\

$\sigma $ & Standard Deviation\\

$\sigma_{\textit{X}^{\hat{}}\textit{y}^{*}} $ & Cross-correlation of the zero mean data X\emph{\textsuperscript{\^{}}} and \emph{y*}\\

\bottomrule
\end{tabularx}}

    \end{table}

\section{Materials and Methods \label{sect:sec2-sensors-1607106}}

\subsection{Support Vector Machine (SVM) \label{sect:sec2dot1-sensors-1607106}}

The SVM is generally used as a classifier that competently categorizes the data using a supervised machine learning algorithm~\cite{B39-sensors-1607106}. It creates an optimal hyperplane using the training data to classify the test data. The optimal hyperplane, also known as support vector, creates a decision border from the nearby samples of different data. If the data are linearly inseparable in their original finite dimensions, then they can be remapped into relatively higher dimensions using kernel tricks. For more details, readers may follow the details described in~\cite{B40-sensors-1607106,B41-sensors-1607106}.

\subsection{Wavelet Packet Decomposition (WPD) \label{sect:sec2dot2-sensors-1607106}}

The wavelet transform decomposes the signals into approximation (cA) and \mbox{detail (cD)} coefficients~\cite{B42-sensors-1607106}---cD consists of high-frequency information about the signal, whereas cA consists of low-frequency information. In DWT, after decomposing the signal into cA and cD, only the cA is being decomposed for further higher-level decomposition. To cover the shortage of fixed time-frequency decomposition in DWT, WPD was proposed as an extension of DWT~\cite{B43-sensors-1607106}. The WPD not only decomposes the cA but also cD simultaneously. As a result, the WPD has the same frequency bandwidth in each resolution while DWT does not.

\textls[-20]{For an EEG signal \emph{x}(\emph{t}), the WPD coefficients can be derived as shown in Equation (1)~\cite{B43-sensors-1607106}:}\begin{equation}
\label{eq:FD1-sensors-1607106}
\left\{ \begin{array}{c}
{d_{0,0}\left( t \right) = x\left( t \right),} \vspace{3pt}\\
{d_{i,2j - 1}\left( t \right) = \sqrt{2}\mathlarger{\mathlarger{\mathlarger{\sum\nolimits_{k}}}}h\left( k \right)d_{i - 1,j}\left( {2t - k} \right),} \vspace{3pt}\\
{d_{i,2j}\left( t \right) = \sqrt{2}\mathlarger{\mathlarger{\mathlarger{\sum\nolimits_{k}}}}g\left( k \right)d_{i - 1,j}\left( {2t - k} \right)} \\
\end{array} \right.
\tag{1}
\end{equation}

\subsection{Wavelet Base and Decomposition Level Selection \label{sect:sec2dot3-sensors-1607106}}

In the wavelet-transform-based signal analysis, the selection of the appropriate mother wavelet and decomposition level is a challenging task. Several mother wavelets are available in the wavelet family. The performance of the wavelet transform is affected by individual wavelet functions. Because hit and trial approach is a time-consuming task, the appropriate mother wavelet is selected according to the reconstruction error in this paper. It is to be noted that when any transformation technique transforms a signal into another domain, and again reconstructs the original signal through inverse operation, the reconstruction error should be ideally zero. The wavelet function is selected that gives the minimum reconstruction error. The reconstruction error ($\varepsilon $) is calculated from Equation (2):\begin{equation}
\label{eq:FD2-sensors-1607106}
\varepsilon = \sqrt{MSE\left( {X - X_{recon}} \right)}~~
\tag{2}
\end{equation}

\textls[-20]{The decomposition level is selected by calculating the Shannon entropy (non-normalized) from the wavelet coefficients~\cite{B44-sensors-1607106}. Non-normalized Shannon entropy is calculated as shown in Equation (3)}:\begin{equation}
\label{eq:FD3-sensors-1607106}
SE_{j} = \mathlarger{\mathlarger{\sum\limits_{K = 1}}}^{N}E_{jk}\log E_{jk}~
\tag{3}
\end{equation}

$E_{jk}~ $ defines as:\begin{equation}
\label{eq:FD4-sensors-1607106}
ED_{jk} = \left| {\text{cD}_{j}\left( k \right)} \right|^{2}~
\tag{4}
\end{equation}
\begin{equation}
\label{eq:FD5-sensors-1607106}
EA_{jk} = \left| {\text{cA}_{j}\left( k \right)} \right|^{2}~~
\tag{5}
\end{equation}

Because most of the information in the wavelet-transformed signals are stored in the approximation coefficients, the optimum level of decomposition is selected at which the entropy of cA becomes less than that of cD.

\subsection{Grey Wolf Optimization (GWO) \label{sect:sec2dot4-sensors-1607106}}

A swarm-based intelligent algorithm, GWO, was first proposed by Mirjalili et~al.~\cite{B45-sensors-1607106}, which is followed by the hunting strategy of grey wolves. GWO is a global search-based optimizer that finds the optimum value through mimicking the hunting strategy of grey wolves~\cite{B46-sensors-1607106}. Wolves are grouped into four classes, for example: alpha, beta, delta, and omega. Alpha is the strongest among the wolves and is followed by beta, delta, and omega. The hunting process follows three major steps: searching for prey, encircling the prey, and finally attacking the prey. For more details readers may refer to~\cite{B45-sensors-1607106}.

\subsection{NLM Algorithm \label{sect:sec2dot5-sensors-1607106}}

The goal of the NLM algorithm is to solve the issues with local smoothing filters by computing the smoothed value as a weighted average of other values in the signal based on the similarity of the neighborhoods~\cite{B47-sensors-1607106}. The weights depend on the similarity between blocks/patch. In the NLM algorithm, the similarity between two blocks (patches) is measured in terms of similarity between their neighborhoods. However, employing the NLM method for denoising of EEG signal is very challenging because of the non-stationary characteristics of the EEG signals.

Let v be noisy observation, where, \emph{u} is the original signal and \emph{n} is the noise, then v is expressed as in Equation (6):\begin{equation}
\label{eq:FD6-sensors-1607106}
v = u + n
\tag{6}
\end{equation}

Then, the estimation $\widetilde{u}\left( s \right) $ for a given sample \emph{s} is calculated by weighted sum of the values at the other point $\eta $ that are within \emph{N}(\emph{s}), where, \emph{N}(\emph{s}) is the “search neighbourhood”, as described in Equations (7) and (8)~\cite{B48-sensors-1607106}.
        \begin{equation}
\label{eq:FD7-sensors-1607106}
\ \widetilde{u}\left( s \right) = \frac{1}{Z\left( s \right)}\mathlarger{\mathlarger{\sum\limits_{\eta\epsilon N{(s)}}}}w\left( {s,\eta} \right)v\left( \eta \right)
\tag{7}
\end{equation}
\begin{equation}
\label{eq:FD8-sensors-1607106}
Z\left( s \right) = \mathlarger{\mathlarger{\sum\nolimits_{\eta}}}w\left( {s,\eta} \right)
\tag{8}
\end{equation}

And the weights \emph{w}(\emph{s},$\ \eta $) are expressed as~\cite{B48-sensors-1607106}:\vspace{-3pt}\begin{equation}
\label{eq:FD9-sensors-1607106}
\begin{array}{c}
{{w\left( {s,\eta} \right) = \exp \left( {- \frac{\sum_{\delta\epsilon\Delta}\left( {v\left( {s + \delta} \right) - v\left( {\eta + \delta} \right)} \right)^{2}}{2L_{\Delta}\lambda^{2}}} \right)}}\vspace{3pt}\\
{{\equiv \exp \left( {- \frac{d^{2}\left( {s,\eta} \right)}{2L_{\Delta}\lambda^{2}}} \right)}}\\
\end{array}
\tag{9}
\end{equation}

It is evident from Equation (9) that the NLM algorithm extracts the non-local information contained in the data. Therefore, the similarity between non-local patches is exploited in the NLM technique. It is to be noted that the weights depend upon the patch similarities instead of the gap between them~\cite{B48-sensors-1607106}. Hence, averaging over similar patches yields a better estimation of the sample. Subsequently, the EEG signal is repetitive in nature and the NLM algorithm may be considered an effective denoising technique.

In the NLM algorithm, estimation is achieved based on the selected similar patches. The main challenge associated with NLM-based EEG denoising is the selection of appropriate filter parameters. There are three key parameters: bandwidth parameter ($\uplambda$), search neighbourhood half-width (M), and patch half-width (P) that need to be optimized for a better estimation. The value of P determines the number of patches. The size of M specifies the area of search, and hence, the larger size of M results in higher computational complexity. Generally, M is limited to reduce the complexity of the system. Out of the three parameters, selection of appropriate $\uplambda$ is more complicated. The level of the signal’s smoothness is determined by the $\uplambda$. A very small value leads to inadequate averaging, while a very large number can induce signal blurring by causing dissimilar patches to look like identical patches~\cite{B49-sensors-1607106}. Generally, $\uplambda$ is calculated as 0.06$\upvarsigma$, $\upvarsigma$\textsuperscript{2} is the variance of noise and equal to the 0.002 microvolts~\cite{B49-sensors-1607106}. However, only the noisy signal is observed; it is more complicated to estimate the standard deviation of the noise before applying any denoising technique to the corrupted EEG signal. Defining a universal value of $\uplambda$ for EEG signal analysis may be unrealistic as the artifacts that contaminate the EEG signals are not always uniform. A modified NLM algorithm is proposed in this paper, in which the optimum value of $\uplambda$ is selected through meta-heuristic algorithm.

\subsection{Performance Matrices \label{sect:sec2dot6-sensors-1607106}}

To validate and test the proposed method, following parameters are evaluated:

\subsubsection{Mutual Information (MI) \label{sect:sec2dot6dot1-sensors-1607106}}

\textls[-30]{The mutual information measures how much information is drawn from the original corrupted EEG signal to reconstructed clean EEG signal~\cite{B50-sensors-1607106}. MI is calculated as in Equation (10):}\begin{equation}
\label{eq:FD11-sensors-1607106}
\text{MI} = \iint\nolimits_{- \infty}^{\infty}P\left( {X,X^{\hat{}}} \right)\log \left( \frac{P\left( {X,X^{\hat{}}} \right)}{P\left( X \right)P\left( X^{\hat{}} \right)} \right)dX~dX^{\hat{}}
\tag{10}
\end{equation}

\subsubsection{Average Correlation Coefficient (CC) \label{sect:sec2dot6dot2-sensors-1607106}}

The average CC computes the degree of similarity between original clean EEG signal and reconstructed clean EEG signal of same duration~\cite{B50-sensors-1607106}. The average CC is calculated as in Equation (11):\begin{equation}
\label{eq:FD12-sensors-1607106}
\text{CC} = \frac{C\left( {X^{\hat{}},y^{*}} \right)}{\sqrt{C\left( {y^{*},y^{*}} \right)~ \ast ~C\left( {X^{\hat{}},X^{\hat{}}} \right)}}~
\tag{11}
\end{equation}

\subsubsection{Structure Similarity Index (SSIM) \label{sect:sec2dot6dot3-sensors-1607106}}

The SSIM calculates the similarity index between original clean EEG signal and reconstructed clean EEG signal of same duration~\cite{B50-sensors-1607106}. It is defined as in Equation (12):\begin{equation}
\label{eq:FD13-sensors-1607106}
\text{SSIM} = \left( \frac{2\mu_{X^{\hat{}}}\mu_{y^{*}}}{\mu_{X^{\hat{}}}^{2} + \mu_{y^{*}}^{2}} \right) \times \left( \frac{2\sigma_{X^{\hat{}}}\sigma_{y^{*}}}{\sigma_{X^{\hat{}}}^{2} + \sigma_{y^{*}}^{2}} \right) \times \left( \frac{\sigma_{X\hat{}y^{*}}}{\sigma_{X^{\hat{}}}\sigma_{y^{*}}} \right)
\tag{12}
\end{equation}

The SSIM and average CC are evaluated on simulated EEG data as the ground truth is available. In the case of recorded real EEG dataset, the ground truth is unavailable, and hence to determine how much information are preserved from corrupted EEG to reconstructed clean EEG during denoising process, MI is evaluated.

\section{Proposed Methodology \label{sect:sec3-sensors-1607106}}

\begin{figure}[H]
      \includegraphics[scale=1]{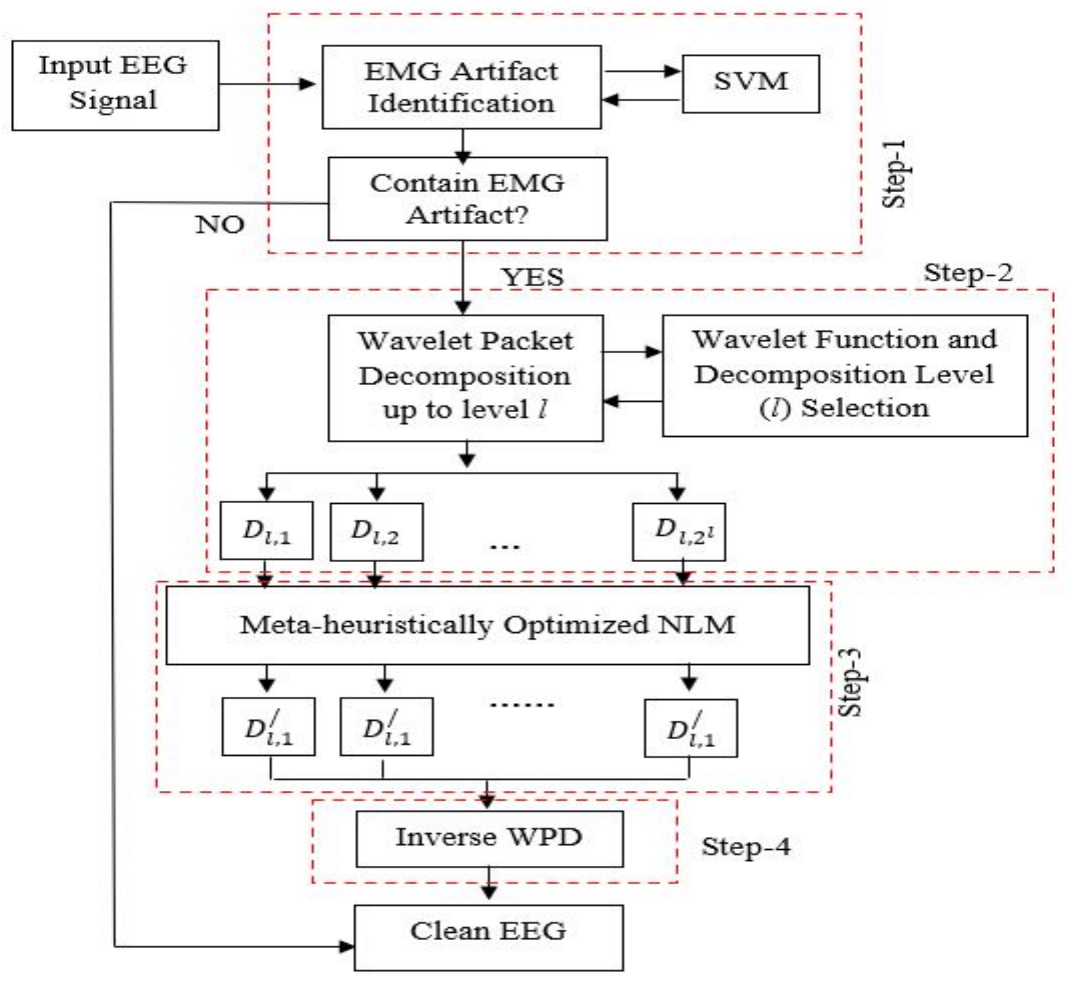}
\caption{The basic block diagram of the proposed algorithm.}
\label{fig:sensors-1607106-f001}
\end{figure}

In this report, a novel hybrid technique for automatic recognition of EMG artifacts and its subsequent elimination from EEG is proposed. The basic flowchart of the proposed algorithm is shown in Figure \ref{fig:sensors-1607106-f001}. Initially the EEG data is recorded and subsequently processed to remove the artifacts. The stages of the proposed model are as follow:\begin{enumerate}[label=,leftmargin=3.1em,labelsep=4mm]
\item[\mbox{Step 1}:] The corrupted EEG (C-EEG) is identified and separated from the non-corrupted EEG (NC-EEG) with the help of a pre-trained SVM.
\item[\mbox{Step 2}:] The corrupted EEG is decomposed into wavelet coefficients through WPD using mother wavelet function up to level \emph{i} (the mother wavelet and the decomposition level are selected as described in Section \ref{sect:sec2dot3-sensors-1607106}).
\item[\mbox{Step 3}:] The wavelet coefficients are then corrected (estimated) through the weighted average of the non-local patches using optimized NLM algorithm. The key parameter ($\uplambda$) of the NLM algorithm is optimized using meta-heuristic optimization technique.
\item[\mbox{Step 4}:] Finally, all the corrected wavelet coefficients are reconstructed using the inverse operation to obtain the clean EEG signal.
\end{enumerate}

The different steps of the proposed approach are elaborated in the subsequent section.
    
\subsection{Data Recording and Processing \label{sect:sec3dot1-sensors-1607106}}

Data were recorded from 40 subjects consisting of 26 male and 14 female students (mean age: 21.5 years) to measure the stress of individuals~\cite{B51-sensors-1607106}. Ethical clearance was obtained from the institutional ethics committee and a consent from was signed by individual subjects. The subjects performed various recognition and arithmetic tasks that were selected to elicit short-term stress in students while their EEG was being recorded. The data were recorded from 32 channels in accordance with the international 10--20 system for placing the electrodes and were sampled at 128 samples per second (1024 Hz internal). The recording was carried out with a wearable 32-channel EMOTIV-EPOC flex gel kit.

\begin{figure}[H]
      \includegraphics[scale=1]{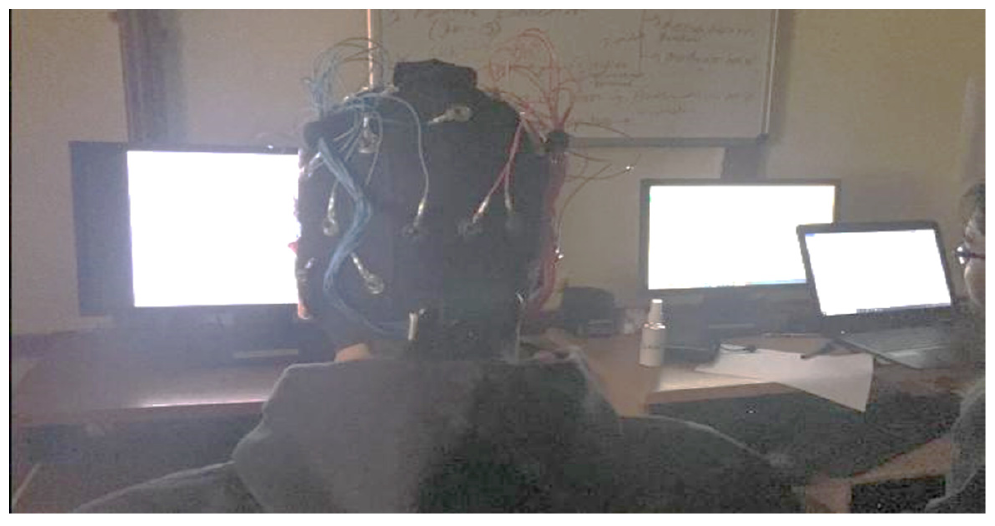}
\caption{Experimental setup: a subject is performing the various mental tasks.}
\label{fig:sensors-1607106-f002}
\end{figure}

The subject was initially asked to relax for a period of 25 s during which relaxing music was played to ease the subject. The subject was instructed to stay calm after completion of every impulse type, after which the instructions for the Stroop color and word test (SCWT) were shown to the subject. The subject was exposed to SCWT for 25 s. The subject then relaxed for 5 s and then the instructions for the next impulse were displayed for 10 s. In the next impulse, the subject was shown mirror images and was asked to identify whether the images are symmetric or asymmetric and respond with a thumbs up or thumbs down gesture, depending on whether the images displayed represent symmetric mirror images or not. The mirror image symmetry task was carried out for 25 s, after which the subject again relaxed for 5 s and then the instructions for the next impulse were displayed for 10 s. Finally, the subject was instructed to solve arithmetic problems mentally and respond with a thumbs up or thumbs down gesture, depending on whether the answer displayed on the screen was a correct solution for the arithmetic problem or not. The arithmetic task was also repeated for 25 s. The completion of the arithmetic task marked the completion of a trial. Moreover, when the subject was responding, an operator also gave feedback as to whether the answers provided by the subject were incorrect or correct. The experimental setup is shown in Figure \ref{fig:sensors-1607106-f002}.    

\vspace{-3pt}

In our experiment, students performed various arithmetic tasks and color tests to measure the stress of individuals. During the experiments it was obvious for a subject to move the head or swallow or move the tongue, which caused the EMG artifacts in the EEG signals. The presence of EMG artifacts in our recorded EEG data were visually inspected by a clinical expert. Generally, normal waveforms of the EEG ranges from 0.1 to 40 Hz. In our experiment, EEG signals were filtered between 0.1 to 40 Hz to remove the frequency components above 40 Hz.

\subsection{Identification of C-EEG \label{sect:sec3dot2-sensors-1607106}}

\begin{figure}[H]
      \includegraphics[scale=0.9]{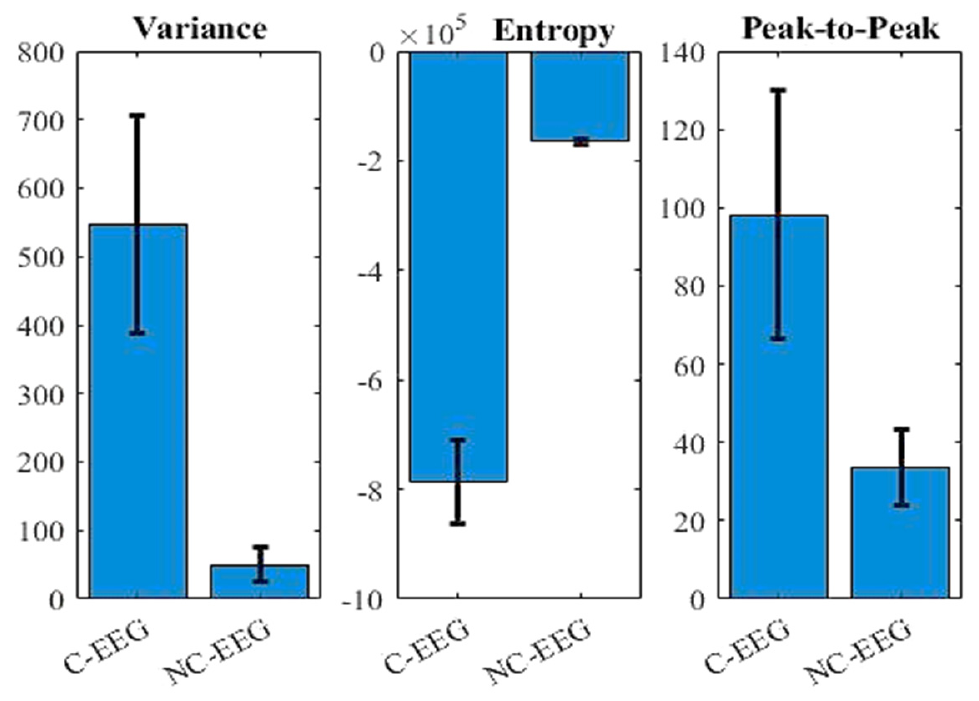}
\caption{Comparison of the average value of features extracted from C$-$EEG and NC$-$EEG.}
\label{fig:sensors-1607106-f003}
\end{figure}

From the recorded data, clean and corrupted EEG segments were extracted by an expert by studying various parameters such as the variance, Shannon entropy. In addition, the expert visually inspected the EEG for segmenting. A total of 400 segments \mbox{(duration: 10 s)} were extracted to represent each of the two classes, namely C-EEG and NC-EEG. The SVM network was trained with the features extracted from both the dataset of same size of 10 s. The data were split in an 80:20 ratio for training and testing. Further, 10-fold cross-validation was used for evaluating the experimental performance. Three distinctive features, variance, Shannon entropy, and peak-to-peak, were calculated from all the signals and provided to train the SVM network. To demonstrate their effectiveness in classifying C-EEG from NC-EEG, an SVM network was trained with these features extracted from EEG data of both the datasets. The average values of these extracted features of NC-EEG, and C-EEG were evaluated respectively and are shown in Figure \ref{fig:sensors-1607106-f003}, which demonstrates the extracted features are distinctive between both the classes. The MATLAB function “fitcsvm” was used to train the classifier with default parameters.    
\vspace{-3pt}

\subsection{Wavelet Decomposition \label{sect:sec3dot3-sensors-1607106}}

After successful identification of the EMG-corrupted EEG signal, it was passed to the second stage of the proposed algorithm for EMG correction. The C-EEG was decomposed into wavelet coefficients through WPD. Wavelet transform was used to separate the EEG features into different scales so that significant features of the EEG signals were preserved and noises could be removed. To cover the shortage of fixed time-frequency decomposition in DWT, WPD was employed in the work, which not only decomposed the cA but also cD simultaneously. As a result, the WPD had the same frequency bandwidth in each resolution while DWT did not.

\subsubsection*{Mother Wavelet and Decomposition Level Selection }

The mother wavelet was carefully chosen by considering the lower reconstruction error as described in Section \ref{sect:sec2dot3-sensors-1607106}. With the minimum reconstruction error, the mother wavelet was selected for decomposing the artifact signal. Several artifact EEG signals were analysed using several mother-wavelet functions, and their average reconstruction error is presented in \ref{tabref:sensors-1607106-t003}. It is observed from the table that “fk6” provides the lowest average reconstruction error in analysing EEG signals. As stated in~\cite{B38-sensors-1607106}, the decomposition level was selected in which the Shannon entropy of cA became less than that of the cD. Hence, the artifact signal was decomposed into wavelet coefficients using “fk6” as mother wavelet up to level 5, and their corresponding Shannon entropy values are compared in \ref{tabref:sensors-1607106-t004}. It is evident from the table that Shannon entropy of cA is less than cD at decomposition \mbox{level 3}. Therefore, in the proposed method, the C-EEG was decomposed into WPD coefficients using “fk6” as mother wavelet up to level 3.    
\vspace{-3pt}
    \begin{table}[H]
    \tablesize{\small}
    \caption{Wavelets vs. average reconstruction error.}

\begin{adjustwidth}{-\extralength}{0cm}
    \label{tabref:sensors-1607106-t003}

\setlength{\cellWidtha}{\fulllength/6-2\tabcolsep-0.5in}
\setlength{\cellWidthb}{\fulllength/6-2\tabcolsep+0.5in}
\setlength{\cellWidthc}{\fulllength/6-2\tabcolsep-0.5in}
\setlength{\cellWidthd}{\fulllength/6-2\tabcolsep+0.5in}
\setlength{\cellWidthe}{\fulllength/6-2\tabcolsep-0.5in}
\setlength{\cellWidthf}{\fulllength/6-2\tabcolsep+0.5in}
\scalebox{1}[1]{\begin{tabularx}{\fulllength}{>{\centering\arraybackslash}m{\cellWidtha}>{\centering\arraybackslash}m{\cellWidthb}>{\centering\arraybackslash}m{\cellWidthc}>{\centering\arraybackslash}m{\cellWidthd}>{\centering\arraybackslash}m{\cellWidthe}>{\centering\arraybackslash}m{\cellWidthf}}
\toprule

\textbf{\textbf{\boldmath{Wavelets}}} & \textbf{\textbf{\boldmath{$\textbf{Average}\textbf{~}{\varepsilon} $}}\textbf{\boldmath{ $\boldmath{\pm}$ Std}}} & \textbf{\textbf{\boldmath{Wavelets}}} & \textbf{$\textbf{Average}\textbf{~}{\varepsilon} $\textbf{\boldmath{ $\boldmath{\pm}$ Std}}} & \textbf{Wavelets} & \textbf{$\textbf{Average}\textbf{~}{\varepsilon} $\textbf{\boldmath{ $\boldmath{\pm}$ Std}}}\\
\cmidrule{1-6}

        Haar
       & 4.07 $\times$ 10\textsuperscript{$-$12} $\pm$ 3.21 $\times$ 10\textsuperscript{$-$12} & db11 & 3.88 $\times$ 10\textsuperscript{$-$12} $\pm$ 1.37 $\times$ 10\textsuperscript{$-$12} & coif2 & 4.45 $\times$ 10\textsuperscript{$-$10} $\pm$ 1.35 $\times$ 10\textsuperscript{$-$10}\\

        db2
       & 2.62 $\times$ 10\textsuperscript{$-$11} $\pm$ 7.07 $\times$ 10\textsuperscript{$-$12} & db12 & 3.81 $\times$ 10\textsuperscript{$-$12} $\pm$ 1.57 $\times$ 10\textsuperscript{$-$12} & coif3 & 2.38 $\times$ 10\textsuperscript{$-$11} $\pm$ 7.33 $\times$ 10\textsuperscript{$-$12}\\

        db3
       & 2.61 $\times$ 10\textsuperscript{$-$10} $\pm$ 9.00 $\times$ 10\textsuperscript{$-$11} & sym2 & 2.62 $\times$ 10\textsuperscript{$-$11} $\pm$ 7.70 $\times$ 10\textsuperscript{$-$12} & coif4 & 1.10 $\times$ 10\textsuperscript{$-$9} $\pm$ 3.12 $\times$ 10\textsuperscript{$-$10}\\

        db4
       & 4.91 $\times$ 10\textsuperscript{$-$11} $\pm$ 1.67 $\times$ 10\textsuperscript{$-$11} & sym3 & 2.61 $\times$ 10\textsuperscript{$-$10} $\pm$ 9.0 $\times$ 10\textsuperscript{$-$11} & coif5 & 2.43 $\times$ 10\textsuperscript{$-$7} $\pm$ 7.01 $\times$ 10\textsuperscript{$-$8}\\

        db5
       & 7.68 $\times$ 10\textsuperscript{$-$11} $\pm$ 2.83 $\times$ 10\textsuperscript{$-$11} & sym4 & 2.29 $\times$ 10\textsuperscript{$-$11} $\pm$ 6.65 $\times$ 10\textsuperscript{$-$12} & fk4 & 8.99 $\times$ 10\textsuperscript{$-$14} $\pm$ 7.67 $\times$ 10\textsuperscript{$-$14}\\

        db6
       & 4.58 $\times$ 10\textsuperscript{$-$11} $\pm$ 1.77 $\times$ 10\textsuperscript{$-$11} & sym5 & 8.08 $\times$ 10\textsuperscript{$-$12} $\pm$ 2.34 $\times$ 10\textsuperscript{$-$12} & \textbf{\boldmath{fk6}} & 5.97 $\times$ 10\textsuperscript{$-$14} $\pm$ 3.26 $\times$ 10\textsuperscript{$-$14}\\

        db7
       & 6.18 $\times$ 10\textsuperscript{$-$11} $\pm$ 1.93 $\times$ 10\textsuperscript{$-$11} & sym6 & 3.57 $\times$ 10\textsuperscript{$-$11} $\pm$ 1.03 $\times$ 10\textsuperscript{$-$11} & fk8 & 7.56 $\times$ 10\textsuperscript{$-$8} $\pm$ 2.19 $\times$ 10\textsuperscript{$-$8}\\

       db8
       & 1.27 $\times$ 10\textsuperscript{$-$10} $\pm$ 4.50 $\times$ 10\textsuperscript{$-$11} & sym7 & 3.47 $\times$ 10\textsuperscript{$-$11} $\pm$ 1.00 $\times$ 10\textsuperscript{$-$11} & fk14 & 1.12 $\times$ 10\textsuperscript{$-$10} $\pm$ 3.26 $\times$ 10\textsuperscript{$-$11}\\

        db9
       & 1.38 $\times$ 10\textsuperscript{$-$9} $\pm$ 4.01 $\times$ 10\textsuperscript{$-$10} & sym8 & 7.01 $\times$ 10\textsuperscript{$-$12} $\pm$ 2.50 $\times$ 10\textsuperscript{$-$12} & fk18 & 7.58 $\times$ 10\textsuperscript{$-$10} $\pm$ 6.69 $\times$ 10\textsuperscript{$-$10}\\

        db10
       & 1.46 $\times$ 10\textsuperscript{$-$10} $\pm$ 4.40 $\times$ 10\textsuperscript{$-$11} & coif1 & 4.23 $\times$ 10\textsuperscript{$-$11} $\pm$ 1.23 $\times$ 10\textsuperscript{$-$11} & fk22 & 1.80 $\times$ 10\textsuperscript{$-$8} $\pm$ 1.18 $\times$ 10\textsuperscript{$-$8}\\

\bottomrule
\end{tabularx}}
\end{adjustwidth}

    \end{table}
    \vspace{-12pt}
    
    \begin{table}[H]
    \tablesize{\small}
    \caption{Shannon entropy value corresponding to the different decomposition levels (mother \mbox{wavelet: “\textbf{\boldmath{fk6}}”).}}
    \label{tabref:sensors-1607106-t004}

\setlength{\cellWidtha}{\textwidth/3-2\tabcolsep-0in}
\setlength{\cellWidthb}{\textwidth/3-2\tabcolsep-0in}
\setlength{\cellWidthc}{\textwidth/3-2\tabcolsep-0in}
\scalebox{1}[1]{\begin{tabularx}{\textwidth}{>{\centering\arraybackslash}m{\cellWidtha}>{\centering\arraybackslash}m{\cellWidthb}>{\centering\arraybackslash}m{\cellWidthc}}
\toprule

\multirow{2.5}{*}{\parbox{\cellWidtha}{\centering \textbf{Decomposition Level}}} & \multicolumn{2}{>{\centering\arraybackslash}m{\cellWidthb + \cellWidthc+2\tabcolsep}}{\textbf{Shannon Entropy}}\\
\cmidrule{2-3}
 & \textbf{DetailCoefficients} & \textbf{Approximation Coefficients}\\
\cmidrule{1-3}

1 & 1.69 $\times$ 10\textsuperscript{4} & 3.07 $\times$ 10\textsuperscript{4}\\

2 & 1.30 $\times$ 10\textsuperscript{3} & 1.69 $\times$ 10\textsuperscript{2}\\

3 & 0.98 $\times$ 10\textsuperscript{3} & $-$1.32 $\times$ 10\textsuperscript{2}\\

4 & 786.31 & $-$1.68 $\times$ 10\textsuperscript{3}\\

5 & $-$409.13 & $-$3.08 $\times$ 10\textsuperscript{4}\\

\bottomrule
\end{tabularx}}

    \end{table}

\subsection{Correction of the Wavelet Coefficients \label{sect:sec3dot4-sensors-1607106}}

It is to be noted that the EMG artifacts present in the EEG signal will be reflected in the corresponding wavelet coefficients. Hence, correcting them will successfully remove the EMG artifacts from the EEG signal. Now the artifact free wavelet coefficients are estimated from the artifact wavelet coefficients using weighted average of the similar patches as described in Equation (7). The estimation is performed through the NLM algorithm. The selection of the optimum parameter of the NLM algorithm is described below.

\begin{figure}[H]
      \includegraphics[scale=1.2]{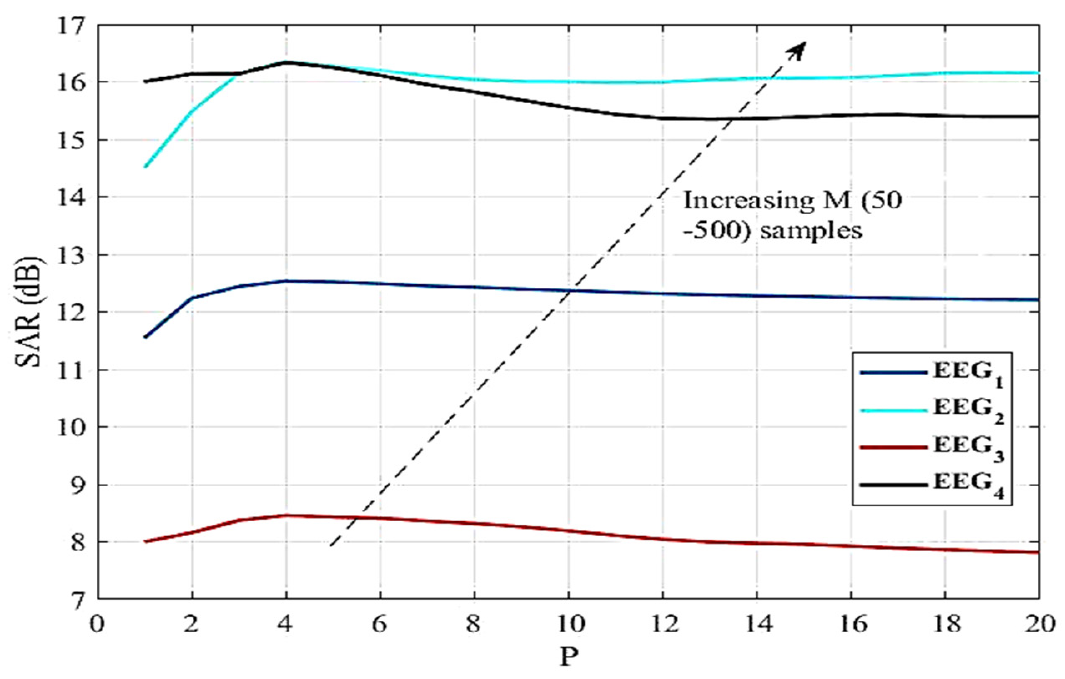}
\caption{Demonstration of SAR improvement for different values of P at $\uplambda$ = 0.0469. Curves are plotted for different M (50--500). The arrow denotes the increasing values of M.}
\label{fig:sensors-1607106-f004}
\end{figure}

\begin{figure}[H]
      \includegraphics[scale=1]{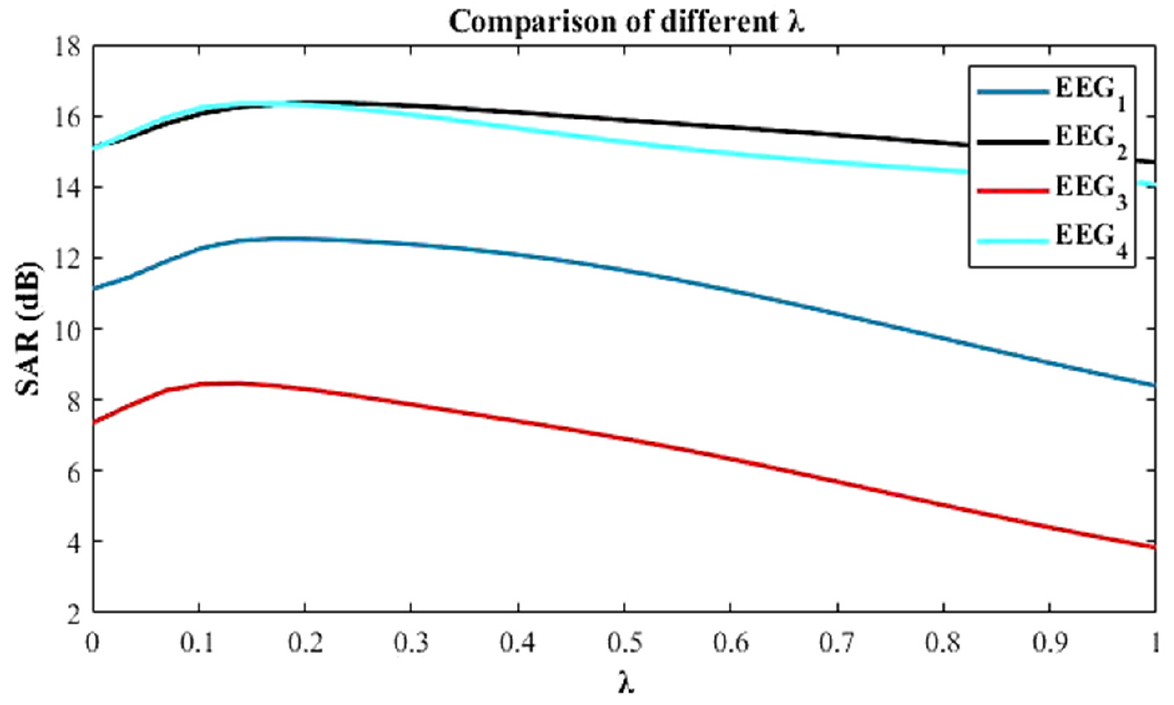}
\caption{SAR improvements for different values of $\uplambda$ at P = 4 and M = 50.}
\label{fig:sensors-1607106-f005}
\end{figure}

An artifact EEG signal was decomposed into WPD coefficients. All the artifacts will be reflected in WPD coefficients. Then, artifact-free coefficients were estimated through the NLM algorithm by tuning its parameters (P, M, $\uplambda$). Four corrupted EEG signals of duration 10 s were analysed by varying the parameters of the NLM algorithm. The SAR improvements achieved by varying several values of P and M are shown in Figure \ref{fig:sensors-1607106-f004}. It is evident from the figure that SAR varies with the different values of P. However, P = 4 provides the maximum SAR for different EEG signals. It is also noted that M does not affect the SAR, but it increases the computational time as shown in \ref{tabref:sensors-1607106-t005}. The SAR values for different $\uplambda$ are also compared and shown in Figure \ref{fig:sensors-1607106-f005}. Unlike P and M, $\uplambda$ is different for different EEG signals yielding the highest SAR. Hence, for NLM-based automatic EEG denoising, it is required to optimize the $\uplambda$ to prevent the over smoothing and incorrect patch similarity. In addition, wavelet coefficients at different scales have different characteristics; hence, single $\uplambda$ for all the scales will degrade the smoothness of the EEG signal. In the proposed work, 2\emph{\textsuperscript{i}} number of $\uplambda$ were optimized for the 2\emph{\textsuperscript{i}} number of wavelet coefficients (where, \emph{i} is the decomposition level) through GWO. The selection of appropriate parameters of the NLM algorithm is explained in Section \ref{sect:sec3dot4dot1-sensors-1607106}.   
 \vspace{-6pt}
\vspace{-9pt}
    
    \begin{table}[H]
    \tablesize{\small}
    \caption{Comparison of the computation time for different values of M.}
    \label{tabref:sensors-1607106-t005}

\setlength{\cellWidtha}{\textwidth/2-2\tabcolsep-0in}
\setlength{\cellWidthb}{\textwidth/2-2\tabcolsep-0in}
\scalebox{1}[1]{\begin{tabularx}{\textwidth}{>{\centering\arraybackslash}m{\cellWidtha}>{\centering\arraybackslash}m{\cellWidthb}}
\toprule

\textbf{M (Samples)} & \textbf{Computational Time (s)}\\
\cmidrule{1-2}

M = 50 & 1.8670\\

M = 100 & 2.7999\\

M = 150 & 3.7531\\

M = 200 & 4.6373\\

M = 250 & 5.4545\\

M = 300 & 6.5207\\

M = 350 & 7.2739\\

M = 400 & 8.0866\\

M = 450 & 8.8712\\

M = 500 & 10.4860\\

\bottomrule
\end{tabularx}}

    \end{table}
    \vspace{-6pt}
    
\subsubsection{Parameter Selection of NLM-Based Denoising \label{sect:sec3dot4dot1-sensors-1607106}}

The parameters of the NLM algorithm were carefully chosen so that the proposed algorithm gave the best performance for removing EMG artifacts from the EEG signals. In the proposed algorithm, several sizes of P and M were investigated and their performance in terms of SAR was compared in the above mentioned Figure \ref{fig:sensors-1607106-f004}. It is observed from the figure that the P = 4 performed better in terms of SAR and M = 50 for lower computational time irrespective of the characteristics of EEG signal.

The bandwidth parameter, $\uplambda$ is the most important parameter in the NLM algorithm as it decides the degree of smoothness. In the proposed algorithm, $\uplambda$ was selected through meta-heuristic optimization technique as to avoid the over-smoothing problem as well as to select the correct similar patches. Two meta-heuristic optimizers, PSO, widely used by researchers due to its fast convergence, and GWO, due to its better execution, were employed for searching the optimal $\uplambda$ for performance comparison.

Let \emph{x}(\emph{t}) be an EEG signal and \emph{n}(\emph{t}) be a muscle artifact; together they formed C-EEG as in Equation (13):\begin{equation}
\label{eq:FD14-sensors-1607106}
Y = x\left( t \right) + n\left( t \right)
\tag{13}
\end{equation}

Now the WPD coefficient $d_{i,j} $ is extracted from the \emph{Y} using “fk6” as mother wavelet up to level \emph{i}. Next, corrected WPD coefficients ($d_{i,j}^{\hat{}} $) were estimated from $d_{i,j} $ using the proposed optimized NLM algorithm. The bandwidth parameter $\text{$\uplambda$}_{n}~\left( {n = 1,2,\ldots,2^{i}} \right) $ was optimized for each coefficient vector. For example, in the proposed method, \emph{Y} was decomposed up to level 3, which implies that eight coefficient vectors were extracted from \emph{Y} as $D_{3,1},~D_{3,2},~\ldots,~D_{3,8} $. To remove the EMG artifacts from each coefficient vectors, eight values of $\uplambda$ were optimized in the proposed method. Mathematically $d_{i,j}^{\hat{}} $ was estimated as in Equation (14):\begin{equation}
\label{eq:FD15-sensors-1607106}
d_{i,j}^{\hat{}} = f\left( {d_{i,j},~P,~M,~\text{$\uplambda$}_{n}~} \right)~~
\tag{14}
\end{equation}
          where \emph{f}(.) denotes the NLM function. As suggested in~\cite{B44-sensors-1607106}, the bandwidth parameter $\uplambda$ was selected as 0.06$\upvarsigma$, where $\upvarsigma$\textsuperscript{2} is the noise-variance and equal to the 0.002. However, EMG artifacts vary with muscle contraction; in this case, $\upvarsigma$\textsuperscript{2} may vary. Considering this fact, $\uplambda$ was randomly initialized between 0.01 to 0.9 in the proposed algorithm. The fitness function used for PSO and GWO is stated in Equations (15) and (16):\begin{equation}
\label{eq:FD16-sensors-1607106}
fitness~function = max\left( {SAR} \right)
\tag{15}
\end{equation}
\begin{equation}
\label{eq:FD17-sensors-1607106}
SAR_{j} = 10\log _{10}\frac{std\left( d_{i,j} \right)}{std( {d_{i,j} - d_{i,j}^{\hat{}}} )}~
\tag{16}
\end{equation}

\subsubsection{Correction \label{sect:sec3dot4dot2-sensors-1607106}}

After estimating the parameters of the NLM algorithm, the corrupted wavelet coefficients were corrected through the modified NLM algorithm. In the proposed method, P = 4 and M = 50 was selected through a hit-and-trial approach, and the $\uplambda$ was selected through GWO. Next, all the corrected wavelet coefficients were passed to the final stage of the proposed method.

\subsection{Reconstruction \label{sect:sec3dot5-sensors-1607106}}

After successfully estimating the clean wavelet coefficients, all the corrected coefficients were used in inverse WPD to reconstruct the original artifact-free EEG signal. The inverse WPD function is expressed as Equation (17). Finally, the artifact-free reconstructed clean EEG signal was provided by the proposed model as output.
        \begin{equation}
\label{eq:FD18-sensors-1607106}
X^{\hat{}} = W^{- 1}( d_{i,j}^{\hat{}} )~~
\tag{17}
\end{equation}

\section{Results \label{sect:sec4-sensors-1607106}}

The proposed technique was first tested on simulated EMG artifact EEG data and then verified on multi-channel real EEG data. The testing was performed on a simulated signal as the original clean EEG signal (the ground truth) was known. As a result, it was possible to compare the original EEG signal and its correction after adding EMG artifacts with the original EEG. In this way, using a simulated EEG signal, the proposed method was tested.

\subsection{Test on Single-Channel Simulated EEG \label{sect:sec4dot1-sensors-1607106}}

The EEG signal was contaminated with EMG artifacts, but the information suppressed by the artifacts was not known (there was no ground truth). Hence, to validate the proposed approach, it was evaluated on simulated C-EEG data for which the ground truth was available. The EMG-contaminated C-EEG signal was simulated in MATLAB using the method described in~\cite{B52-sensors-1607106}. At first, the clean EEG segments of duration 10 s were simulated by adding 20 sinusoids together with frequencies selected randomly from 0.1 to 30 Hz with sampling frequency as 250 Hz. Next, the muscle activities were modelled using random noise bandpass filtered between 5 to 45 Hz. Finally, clean EEG segments and muscle activities were added together to form the C-EEG signal.

\begin{figure}[H]
      \includegraphics[scale=1.2]{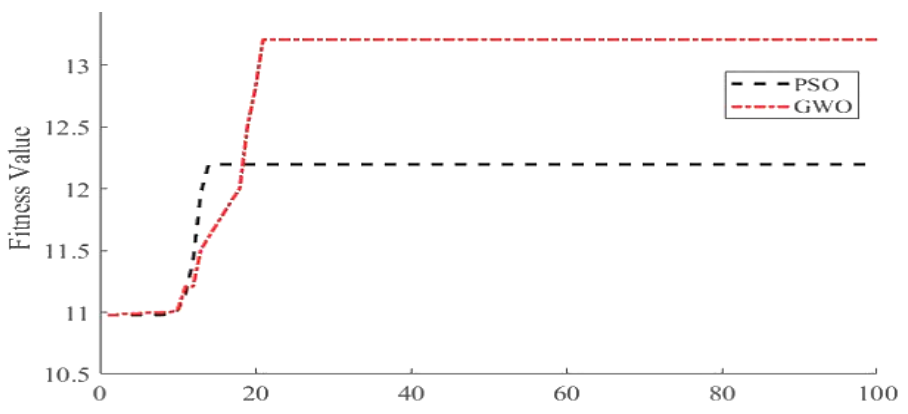}
\caption{Convergence comparison between PSO and GWO.}
\label{fig:sensors-1607106-f006}
\end{figure}

\begin{figure}[H]
          \begin{adjustwidth}{-\extralength}{0cm}
      \centering
      \includegraphics[scale=1]{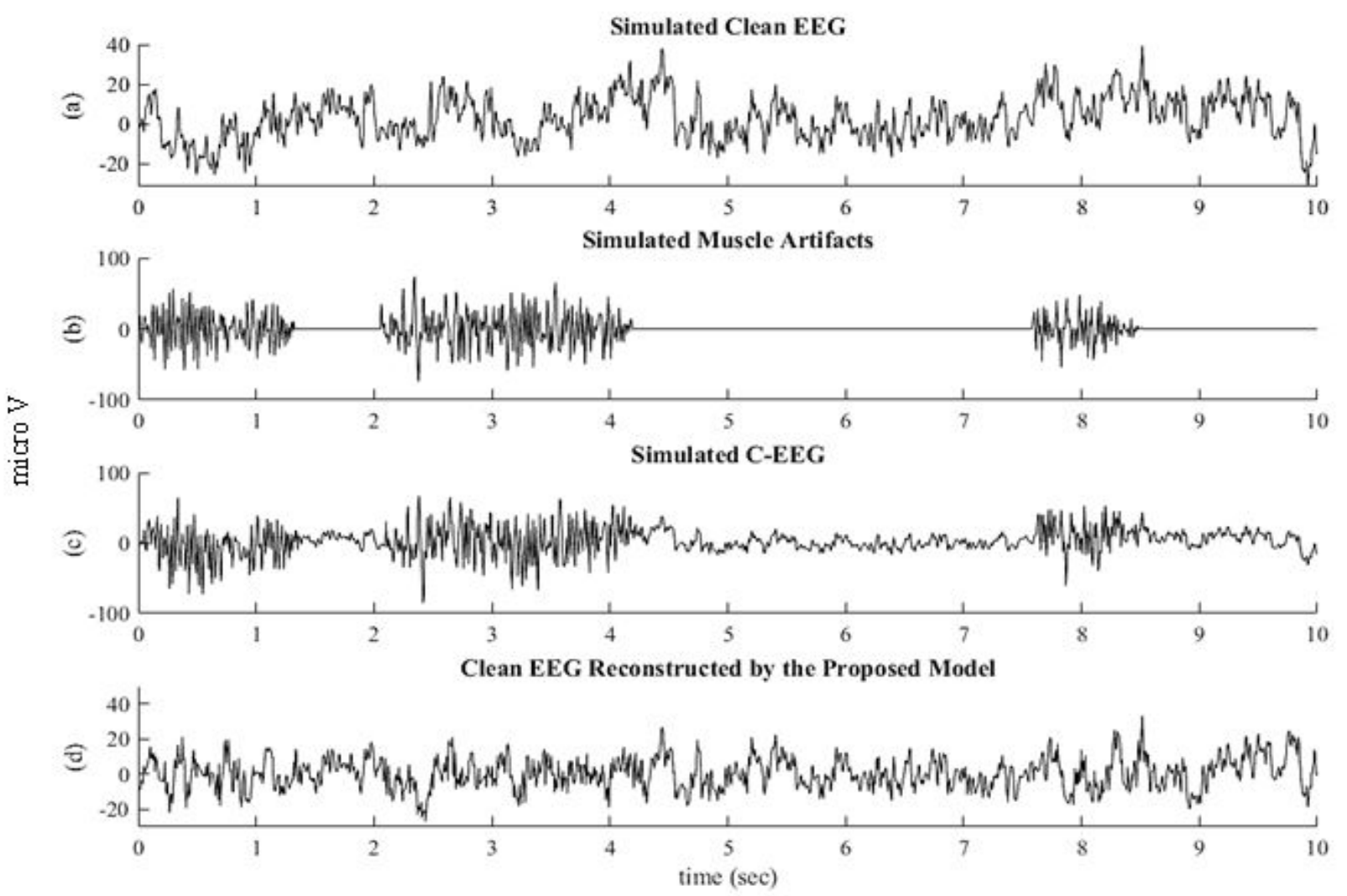}
          \end{adjustwidth}
\caption{Simulated EEG signal and its corrected EEG signal: (\textbf{\boldmath{a}}) Simulated Clean EEG Signal, \mbox{(\textbf{\boldmath{b}}) Simulated} EMG Artifact Signal, (\textbf{\boldmath{c}}) Simulated Corrupted EEG Signal, and (\textbf{\boldmath{d}}) Clean EEG Reconstructed by the proposed model.}
\label{fig:sensors-1607106-f007}
\end{figure}

\begin{figure}[H]
      \includegraphics[scale=0.95]{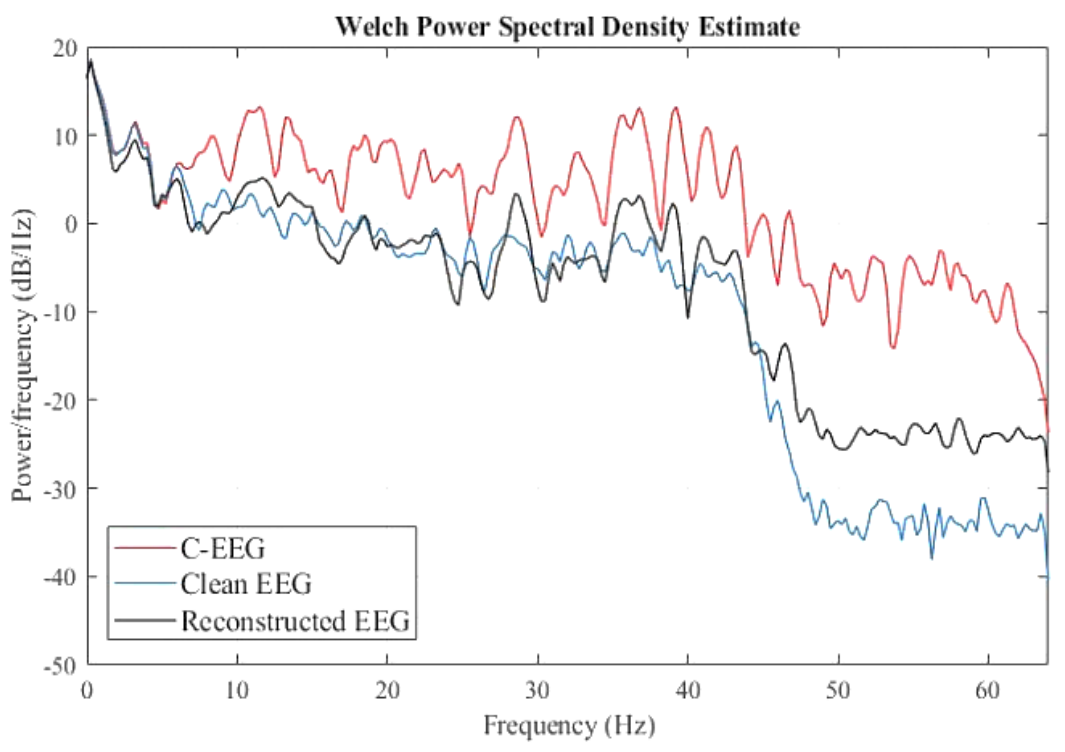}
\caption{Comparison of power spectral density among C$-$EEG, Clean EEG, and Reconstructed EEG.}
\label{fig:sensors-1607106-f008}
\end{figure}

The simulated artifactual EEG signal was passed through the pre-trained SVM as a classifier. The SVM automatically recognized C-EEGs through the extracted features. The performance in terms of accuracy, specificity, and sensitivity of the SVM classifier was compared with the Naïve Bias Classifier (NBC) and is shown in \ref{tabref:sensors-1607106-t006}. It is evident from the table that the SVM effectively classified the C-EEG from NC-EEG using extracted features. Next, the identified C-EEG signal was decomposed into eight WPD coefficient vectors using “fk6” as wavelet function as described above. For estimating clean WPD coefficient vectors, the proposed NLM algorithm was applied on eight coefficient vectors. For finding the optimum $\uplambda$, both the optimizer PSO and GWO were implemented, and their performance in term of convergence is presented in Figure \ref{fig:sensors-1607106-f006}. It is observed from the figure that GWO attains excellence solution with maximum fitness as compared to PSO. The simulated clean EEG signal and EMG signal are shown in Figure \ref{fig:sensors-1607106-f007}a,b, respectively. By adding them, the C-EEG signal was generated and is shown in Figure \ref{fig:sensors-1607106-f007}c. EMG artifacts were observed in the time of 0 to 1.3 s, 2 to 4.2 s, and 7.6 to 8.5 s in C-EEG. In Figure \ref{fig:sensors-1607106-f007}d, the corrected EEG signal denoised through the proposed algorithm is shown. It is observed that the proposed algorithm successfully removed the EMG artifacts from C-EEG while preserving true EEG structure. The power spectral density of C-EEG, clean EEG, and reconstructed EEG were also analysed to verify the proposed algorithm and are compared in Figure \ref{fig:sensors-1607106-f008}. It is observed that the proposed denoising algorithm preserved most of the frequency bands in the reconstructed EEG.    
    \begin{table}[H]
    \tablesize{\small}
    \caption{Performance of the classifier.}
    \label{tabref:sensors-1607106-t006}

\setlength{\cellWidtha}{\textwidth/3-2\tabcolsep-0in}
\setlength{\cellWidthb}{\textwidth/3-2\tabcolsep-0in}
\setlength{\cellWidthc}{\textwidth/3-2\tabcolsep-0in}
\scalebox{1}[1]{\begin{tabularx}{\textwidth}{>{\centering\arraybackslash}m{\cellWidtha}>{\centering\arraybackslash}m{\cellWidthb}>{\centering\arraybackslash}m{\cellWidthc}}
\toprule

\multirow{2.5}{*}{\parbox{\cellWidtha}{\centering \textbf{Performance}}} & \multicolumn{2}{>{\centering\arraybackslash}m{\cellWidthb + \cellWidthc+2\tabcolsep}}{\textbf{Classifiers }}\\
\cmidrule{2-3}
 & \textbf{SVM} & \textbf{NBC}\\
\cmidrule{1-3}

\textbf{\boldmath{Sensitivity}} & 98.31 & 94.09\\

\textbf{\boldmath{Specificity}} & 97.91 & 91.68\\

\textbf{\boldmath{Accuracy}} & 98.19 & 93.28\\

\bottomrule
\end{tabularx}}

    \end{table}
    \vspace{-12pt}
    
\vspace{-9pt}
    
\vspace{-6pt}
    
\begin{figure}[H]
          \begin{adjustwidth}{-\extralength}{0cm}
      \centering
      \includegraphics[scale=1]{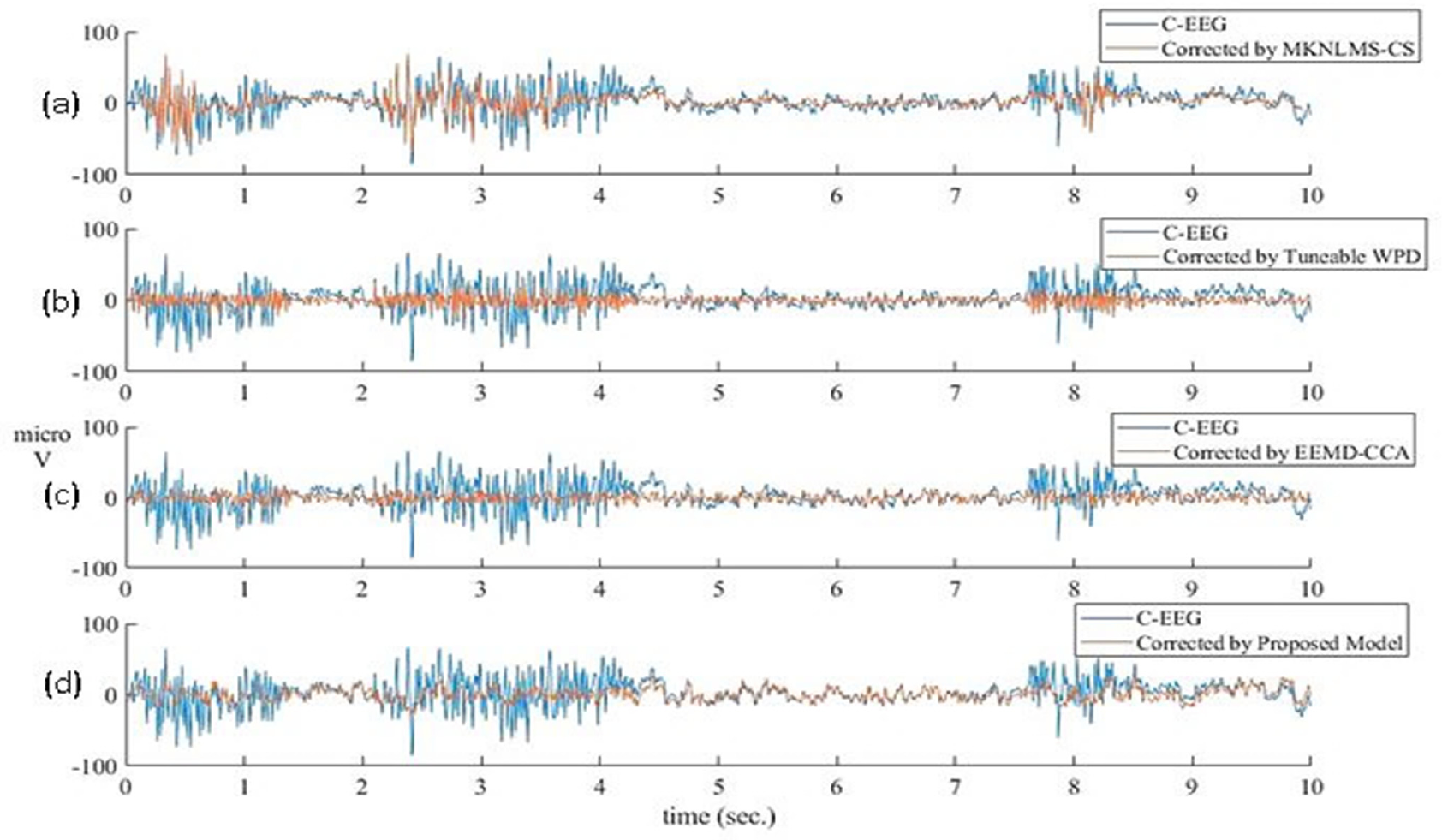}
          \end{adjustwidth}
\caption{Comparison between simulated corrupted EEG and corresponding clean EEG reconstructed by (\textbf{\boldmath{a}}) MKNLMS$-$CS, (\textbf{\boldmath{b}}) Tuneable WPD, (\textbf{\boldmath{c}}) EEMD$-$CCA, and (\textbf{\boldmath{d}}) proposed methodology.}
\label{fig:sensors-1607106-f009}
\end{figure}

To check the superiority, the proposed model was compared with the recently reported EEG denoising techniques~\cite{B14-sensors-1607106,B35-sensors-1607106,B38-sensors-1607106} and is presented in Figure \ref{fig:sensors-1607106-f009}. In Figure \ref{fig:sensors-1607106-f009}a,b, it is observed that MKNLMS-CS~\cite{B38-sensors-1607106} and tuneable WPD~\cite{B35-sensors-1607106} removed muscle artifacts from the C-EEG, respectively, while distorting the other portion of the C-EEG where no muscle artifacts are present. EEMD-CCA~\cite{B14-sensors-1607106} slightly improves the distortion that occurred in the clean portion of the C-EEG as shown in Figure \ref{fig:sensors-1607106-f009}c. In Figure \ref{fig:sensors-1607106-f009}d it is observed that the proposed method successfully removed the muscle artifacts while preserving the maximum information contained in the C-EEG. In addition to the pictorial evidence, how much information was preserved by all the methods during denoising process was also investigated in a quantitative manner. The CC between the clean portion of C-EEG and clean portion of reconstructed clean EEG was evaluated to verify how much information was lost during denoising the EEG. It is observed from Figure \ref{fig:sensors-1607106-f009} that the EEG portion ranged from 4.3 s to 7.3 s and did not contain any muscle artifacts. Hence, CC between the C-EEG portion (ranges from 4.3 s to 7.3 s) and reconstructed clean EEG portion \mbox{(ranges from 4.3 s to 7.3 s)} was calculated with all the denoising techniques and compared in \ref{tabref:sensors-1607106-t007}. It is evident from the table that the proposed model highly preserved the information contained in the original EEG during denoising the EEG signal. Further, CC and SSIM was calculated between C-EEG and the corrected EEG by all the methods. The performance matrices CC and SSIM were evaluated for all four denoising methods and the proposed method, and are compared in \ref{tabref:sensors-1607106-t008}.    
\vspace{-3pt}
   \begin{table}[H]
    \tablesize{\small}
    \caption{\textls[-30]{Comparison of similarity between C-EEG and reconstructed clean EEG (ranges from 4.3 s to 7.3 s).}}
    \label{tabref:sensors-1607106-t007}

\setlength{\cellWidtha}{\textwidth/2-2\tabcolsep-0in}
\setlength{\cellWidthb}{\textwidth/2-2\tabcolsep-0in}
\scalebox{1}[1]{\begin{tabularx}{\textwidth}{>{\centering\arraybackslash}m{\cellWidtha}>{\centering\arraybackslash}m{\cellWidthb}}
\toprule

\textbf{Denoising Techniques} & \textbf{Average CC}\\
\cmidrule{1-2}

\textbf{\boldmath{MKNLMS$-$CS}} {\cite{B38-sensors-1607106}} & 0.2809\\

\textbf{\boldmath{Tuneable WPD}} {\cite{B35-sensors-1607106}} & 0.5089\\

\textbf{\boldmath{EEMD$-$CCA}} {\cite{B14-sensors-1607106}} & 0.5557\\

\textbf{\boldmath{Proposed Method}} & 0.8863\\

\bottomrule
\end{tabularx}}

    \end{table}
    
    \vspace{-12pt}

    \begin{table}[H]
    \tablesize{\small}
    \caption{Performance comparison on simulated C-EEG.}
    \label{tabref:sensors-1607106-t008}

\setlength{\cellWidtha}{\textwidth/3-2\tabcolsep-0in}
\setlength{\cellWidthb}{\textwidth/3-2\tabcolsep-0in}
\setlength{\cellWidthc}{\textwidth/3-2\tabcolsep-0in}
\scalebox{1}[1]{\begin{tabularx}{\textwidth}{>{\centering\arraybackslash}m{\cellWidtha}>{\centering\arraybackslash}m{\cellWidthb}>{\centering\arraybackslash}m{\cellWidthc}}
\toprule

\textbf{Methods} & \textbf{Average CC} & \textbf{SSIM}\\
\cmidrule{1-3}

\textbf{\boldmath{MKNLMS}}\textbf{\boldmath{$-$}}\textbf{\boldmath{CS}} {\cite{B38-sensors-1607106}} & 0.5937 & 0.3963\\

\textbf{\boldmath{Tuneable WPD}} {\cite{B35-sensors-1607106}} & 0.7101 & 0.5401\\

\textbf{\boldmath{EEMD}}\textbf{\boldmath{$-$}}\textbf{\boldmath{CCA}} {\cite{B14-sensors-1607106}} & 0.8139 & 0.5723\\

\textbf{\boldmath{Proposed Method}} & 0.8675 & 0.6809\\

\bottomrule
\end{tabularx}}

    \end{table}
    
    \vspace{-6pt}
    
\subsection{Test on Multi-Channel Recorded EEG Data \label{sect:sec4dot2-sensors-1607106}}

After successfully verifying on single-channel simulated C$-$EEG, the proposed model was also tested on recorded EEG. The 32-channel raw EEG data of duration of 10 s were chosen from the dataset and filtered between 0.1--40 Hz through a bandpass filter. However, all the channels were affected by the EMG artifacts. The proposed model automatically recognized the EMG artifact EEG data and subsequently corrected them. The proposed model successfully eliminated all the EMG artifacts from the corrupted EEG while preserving the true EEG structure. For quantitative analysis, MI between the corrupted EEG signal and reconstructed clean EEG was measured and compared with the recently developed techniques as presented in \ref{tabref:sensors-1607106-t009}.    
\vspace{-3pt}
    \begin{table}[H]
    \tablesize{\footnotesize}
    \caption{Comparison of MI on 32-channel recorded real EEG data.}
    \label{tabref:sensors-1607106-t009}

\setlength{\cellWidtha}{\textwidth/5-2\tabcolsep-0.15in}
\setlength{\cellWidthb}{\textwidth/5-2\tabcolsep+0.1in}
\setlength{\cellWidthc}{\textwidth/5-2\tabcolsep+0.05in}
\setlength{\cellWidthd}{\textwidth/5-2\tabcolsep-0in}
\setlength{\cellWidthe}{\textwidth/5-2\tabcolsep-0in}
\scalebox{1}[1]{\begin{tabularx}{\textwidth}{>{\centering\arraybackslash}m{\cellWidtha}>{\centering\arraybackslash}m{\cellWidthb}>{\centering\arraybackslash}m{\cellWidthc}>{\centering\arraybackslash}m{\cellWidthd}>{\centering\arraybackslash}m{\cellWidthe}}
\toprule

\textbf{Channels} & \textbf{MKNLMS$\boldmath{-}$CS {\cite{B38-sensors-1607106}}} & \textbf{Tuneable WPD {\cite{B35-sensors-1607106}}} & \textbf{EEMD$\boldmath{-}$CCA {\cite{B14-sensors-1607106}}} & \textbf{Proposed Method}\\
\cmidrule{1-5}

\textbf{\boldmath{1}} & 1.009 & 1.106 & 2.417 & 2.970\\
\cmidrule{1-5}
\textbf{\boldmath{2}} & 0.952 & 1.193 & 2.334 & 2.972\\
\cmidrule{1-5}
\textbf{\boldmath{3}} & 1.562 & 1.535 & 2.858 & 3.985\\
\cmidrule{1-5}
\textbf{\boldmath{4}} & 2.389 & 1.986 & 3.198 & 3.908\\
\cmidrule{1-5}
\textbf{\boldmath{5}} & 1.410 & 1.444 & 2.583 & 3.207\\
\cmidrule{1-5}
\textbf{\boldmath{6}} & 1.053 & 1.094 & 2.117 & 2.652\\
\cmidrule{1-5}
\textbf{\boldmath{7}} & 1.119 & 1.274 & 2.426 & 2.958\\
\cmidrule{1-5}
\textbf{\boldmath{8}} & 1.630 & 1.631 & 2.994 & 4.179\\
\cmidrule{1-5}
\textbf{\boldmath{9}} & 1.930 & 1.884 & 2.319 & 3.010\\
\cmidrule{1-5}
\textbf{\boldmath{10}} & 1.427 & 1.361 & 3.202 & 3.359\\
\cmidrule{1-5}
\textbf{\boldmath{11}} & 1.601 & 1.609 & 2.197 & 2.387\\
\cmidrule{1-5}
\textbf{\boldmath{12}} & 1.198 & 1.185 & 2.429 & 2.798\\

\bottomrule
\end{tabularx}}
\end{table}
\begin{table}[H]\ContinuedFloat
 \tablesize{\footnotesize}
\caption{\textit{Cont}.}
  \label{tabref:sensors-1607106-t009}

\setlength{\cellWidtha}{\textwidth/5-2\tabcolsep-0.15in}
\setlength{\cellWidthb}{\textwidth/5-2\tabcolsep+0.1in}
\setlength{\cellWidthc}{\textwidth/5-2\tabcolsep+0.05in}
\setlength{\cellWidthd}{\textwidth/5-2\tabcolsep-0in}
\setlength{\cellWidthe}{\textwidth/5-2\tabcolsep-0in}
\scalebox{1}[1]{\begin{tabularx}{\textwidth}{>{\centering\arraybackslash}m{\cellWidtha}>{\centering\arraybackslash}m{\cellWidthb}>{\centering\arraybackslash}m{\cellWidthc}>{\centering\arraybackslash}m{\cellWidthd}>{\centering\arraybackslash}m{\cellWidthe}}
\toprule

\textbf{Channels} & \textbf{MKNLMS$\boldmath{-}$CS {\cite{B38-sensors-1607106}}} & \textbf{Tuneable WPD {\cite{B35-sensors-1607106}}} & \textbf{EEMD$\boldmath{-}$CCA {\cite{B14-sensors-1607106}}} & \textbf{Proposed Method}\\
\cmidrule{1-5}
\textbf{\boldmath{13}} & 1.039 & 1.165 & 2.335 & 2.842\\
\cmidrule{1-5}
\textbf{\boldmath{14}} & 0.932 & 1.506 & 2.026 & 2.127\\
\cmidrule{1-5}
\textbf{\boldmath{15}} & 1.109 & 1.158 & 2.058 & 2.199\\
\cmidrule{1-5}
\textbf{\boldmath{16}} & 2.011 & 1.054 & 2.042 & 2.292\\
\cmidrule{1-5}
\textbf{\boldmath{17}} & 1.520 & 1.055 & 2.371 & 2.747\\
\cmidrule{1-5}
\textbf{\boldmath{18}} & 1.106 & 1.158 & 2.007 & 2.446\\
\cmidrule{1-5}
\textbf{\boldmath{19}} & 1.321 & 1.033 & 2.063 & 2.496\\
\cmidrule{1-5}
\textbf{\boldmath{20}} & 1.030 & 1.146 & 1.891 & 2.305\\
\cmidrule{1-5}
\textbf{\boldmath{21}} & 1.098 & 0.997 & 2.248 & 2.451\\
\cmidrule{1-5}
\textbf{\boldmath{22}} & 1.032 & 0.900 & 2.120 & 2.421\\
\cmidrule{1-5}
\textbf{\boldmath{23}} & 1.001 & 1.045 & 2.267 & 2.691\\
\cmidrule{1-5}
\textbf{\boldmath{24}} & 1.030 & 1.078 & 2.010 & 1.985\\
\cmidrule{1-5}
\textbf{\boldmath{25}} & 1.131 & 1.132 & 1.864 & 1.783\\
\cmidrule{1-5}
\textbf{\boldmath{26}} & 1.095 & 1.385 & 2.564 & 3.289\\
\cmidrule{1-5}
\textbf{\boldmath{27}} & 1.521 & 1.611 & 3.337 & 3.914\\
\cmidrule{1-5}
\textbf{\boldmath{28}} & 1.302 & 1.535 & 2.930 & 3.181\\
\cmidrule{1-5}
\textbf{\boldmath{29}} & 1.015 & 1.268 & 2.573 & 3.151\\
\cmidrule{1-5}
\textbf{\boldmath{30}} & 1.400 & 1.440 & 2.791 & 3.465\\
\cmidrule{1-5}
\textbf{\boldmath{31}} & 1.612 & 1.615 & 3.082 & 4.607\\
\cmidrule{1-5}
\textbf{\boldmath{32}} & 1.590 & 1.703 & 2.998 & 4.211\\
\cmidrule{1-5}
\textbf{\boldmath{Average $\pm$ Std}} & \textbf{\boldmath{1.3180 $\pm$ 0.3478}} & \textbf{\boldmath{1.3214 $\pm$ 0.2747}} & \textbf{\boldmath{2.4578 $\pm$ 0.4239}} & \textbf{\boldmath{2.9684 $\pm$ 0.7045}}\\

\bottomrule
\end{tabularx}}

    \end{table}

\section{Discussion \label{sect:sec5-sensors-1607106}}

In this paper, a fully automatic denoising method is proposed for removing EMG artifacts from the EEG signals. At first, the artifact EEG signals are identified through SVM as a classifier. After that, the corrupted EEG signals are decomposed into wavelet coefficients through WPD. Wavelet transform is used to separate the EEG features into different scales so that significant features of the EEG signals are preserved and artifacts can be removed. Generally, in wavelet denoising techniques, wavelet coefficients are thresholded to remove the artifacts. However, selection of an appropriate threshold makes the system unfit for automatic operation. In the proposed method, the corrupted wavelet coefficients are corrected through a modified NLM algorithm instead of thresholding them. The NLM algorithm has various parameters and needed adjustment for different types of EEG data. The NLM algorithm has been modified in this paper by optimizing the bandwidth parameter through GWO. In addition, a set of $\uplambda$ is optimized at different scales of the wavelet-transformed EEG signal, which enhances the performance of the proposed method for denoising the EEG signals. Finally, all the corrected wavelet coefficients are used in inverse operation to get back the original clean EEG signal. The proposed method is fully automatic and does not need any intervention of the user.

The proposed method was first validated on simulated EEG signals (the ground truth is available) and then tested on recorded EEG data. It is observed from the experimental results that the proposed method is superior among all the denoising methods in terms of highest CC and SSIM. Hence, the proposed algorithm preserves more true structure of the clean EEG signal compared to other recently reported EEG denoising algorithms. For recorded EEG data, the original true clean EEG is unavailable (there is no ground truth); hence, MI is calculated to check how much information is mutual between corrupted EEG and denoised EEG. It is observed that the proposed approach achieved higher MI as compared to other recently developed techniques.

One of the main advantages of the proposed hybrid method is that it is insensitive to any threshold value and there is no need to tune the parameters of the NLM algorithm, which makes the system fully automatic and it can be implemented in online applications. WPD efficiently deals with the non-stationary characteristics of EEG signals. Although the proposed method is capable of correcting EMG artifacts from the multi-channel EEG signals, correcting one by one increases the computational time.

\section{Conclusions and Future Scope \label{sect:sec6-sensors-1607106}}

In this paper, a new automatic hybrid system for denoising muscle artifacts from EEG is introduced for the first time, in which WPD is combined with an optimized NLM algorithm. The WPD is adopted for its multi-resolution analysis while using only a single-channel EEG. Unlike other current state-of-art methods, the proposed system removes artifacts from its time-frequency response of the EEG signal through an optimized NLM algorithm instead of thresholding. The proposed system removes the muscle artifacts from the EEG signal, no matter how many artifacts contaminate the EEG. Further, several challenges using the NLM algorithm for EEG denoising are discussed and properly addressed in this work. One major challenge associated with the NLM algorithm is the selection of proper bandwidth parameter. It is also shown that the bandwidth parameter is different for different EEG. In this work, a meta-heuristically optimized NLM algorithm is presented, in which the bandwidth parameter is selected according to the nature of the EEG. The proposed system can operate automatically and does not require any intervention of the user while using it. The proposed system was first validated with a simulated C-EEG in which ground truth was available, which demonstrated better performance, and then tested with recorded real EEG data for its practical validation. In addition to a single channel, the proposed system is also able to remove the muscle artifacts from multi-channel EEG data. Future studies are likely to investigate the performance of the proposed method in removing other kinds \mbox{of artifacts.}

\vspace{6pt}
\authorcontributions{Conceptualization, S.P. and E.G.; data curation, R.G.; formal analysis, S.P. and R.G.; investigation, S.P. and N.S.; project administration, N.S.; supervision, N.S.; validation, S.P. and N.S.; visualization, N.S. and R.G.; writing---original draft, S.P.; sriting---review and editing, N.S., R.G. and E.G. All authors have read and agreed to the published version of the manuscript.}
\funding{The research work is funded by NPIU-MHRD, Government of India, under the CRS Project entitled “Monitoring Stress in Students using EEG”, with CRS-ID: 1-5770264050.}
\dataavailability{It is described in the data section and cited the original article~\cite{B51-sensors-1607106}.}
\conflictsofinterest{The authors declare no conflict of interest. The funders had no role in the design of the study; in the collection, analyses, or interpretation of data; in the writing of the manuscript, or in the decision to publish the results.}
\begin{adjustwidth}{-\extralength}{0cm}

\reftitle{References}

\end{adjustwidth}


\begin{thebibliography}{999}
\bibitem{B1-sensors-1607106}
Zhang, X.; Yao, L.; Wang, X.; Monaghan, J.; McAlpine, D.; Zhang, Y. A survey on deep learning-based non-invasive brain signals: Recent advances and new frontiers. \emph{J. Neural Eng.} \textbf{\boldmath{2021}}, \emph{18}, 031002. [\href{http://doi.org/10.1088/1741-2552/abc902}{CrossRef}] [\href{http://www.ncbi.nlm.nih.gov/pubmed/33171452}{PubMed}]

\bibitem{B2-sensors-1607106}
Jin, Z.; Zhou, G.; Gao, D.; Zhang, Y. EEG classification using sparse Bayesian extreme learning machine for brain--computer interface. \emph{Neural Comput. Applic} \textbf{\boldmath{2020}}, \emph{32}, 6601--6609. [\href{http://doi.org/10.1007/s00521-018-3735-3}{CrossRef}]

\bibitem{B3-sensors-1607106}
Ghosh, R.; Sinha, N.; Biswas, S.K. Removal of Eye-Blink Artifact from EEG Using LDA and Pre-trained RBF Neural Network. \mbox{In \emph{Smart} Computing Paradigms: New Progresses and Challenges}; Springer: Singapore, 2020; pp. 217--225.

\bibitem{B4-sensors-1607106}
Phadikar, S.; Sinha, N.; Ghosh, R. A Survey on Feature Extraction Methods for EEG Based Emotion Recognition. In \emph{International Conference on Innovation in Modern Science and Technology}; Springer: Cham, Switzerland, 2019; pp. 31--45.

\bibitem{B5-sensors-1607106}
Shackman, A.J.; McMenamin, B.W.; Slagter, H.A.; Maxwell, J.S.; Greischar, L.L.; Davidson, R.J. Electromyogenic artifacts and electroencephalographic inferences. \emph{Brain Topogr.} \textbf{\boldmath{2009}}, \emph{22}, 7--12. [\href{http://doi.org/10.1007/s10548-009-0079-4}{CrossRef}] [\href{http://www.ncbi.nlm.nih.gov/pubmed/19214730}{PubMed}]

\bibitem{B6-sensors-1607106}
Goncharova, I.I.; McFarland, D.J.; Vaughan, T.M.; Wolpaw, J.R. EMG contamination of EEG: Spectral and topographical characteristics. \emph{Clin. Neurophysiol.} \textbf{\boldmath{2003}}, \emph{114}, 1580--1593. [\href{http://doi.org/10.1016/S1388-2457(03)00093-2}{CrossRef}]

\bibitem{B7-sensors-1607106}
Chen, X.; Liu, Q.; Tao, W.; Li, L.; Lee, S.; Liu, A.; Chen, Q.; Cheng, J.; McKeown, M.J. ReMAE: User-friendly toolbox for removing muscle artifacts from EEG. \emph{IEEE Trans. Instrum. Meas.} \textbf{\boldmath{2019}}, \emph{69}, 2105--2119. [\href{http://doi.org/10.1109/TIM.2019.2920186}{CrossRef}]

\bibitem{B8-sensors-1607106}
Mowla, M.R.; Ng, S.C.; Zilany, M.S.; Paramesran, R. Artifacts-matched blind source separation and wavelet transform for multichannel EEG denoising. \emph{Biomed. Signal Process. Control.} \textbf{\boldmath{2015}}, \emph{22}, 111--118. [\href{http://doi.org/10.1016/j.bspc.2015.06.009}{CrossRef}]

\bibitem{B9-sensors-1607106}
Frølich, L.; Dowding, I. Removal of muscular artifacts in EEG signals: A comparison of linear decomposition methods. \emph{Brain Inform.} \textbf{\boldmath{2018}}, \emph{5}, 13--22. [\href{http://doi.org/10.1007/s40708-017-0074-6}{CrossRef}]

\bibitem{B10-sensors-1607106}
Jiang, X.; Bian, G.B.; Tian, Z. Removal of artifacts from EEG signals: A review. \emph{Sensors} \textbf{\boldmath{2019}}, \emph{19}, 987. [\href{http://doi.org/10.3390/s19050987}{CrossRef}]

\bibitem{B11-sensors-1607106}
Chang, C.Y.; Hsu, S.H.; Pion-Tonachini, L.; Jung, T.P. Evaluation of artifact subspace reconstruction for automatic artifact components removal in multi-channel EEG recordings. \emph{IEEE Trans. Biomed. Eng.} \textbf{\boldmath{2019}}, \emph{67}, 1114--1121. [\href{http://doi.org/10.1109/TBME.2019.2930186}{CrossRef}]

\bibitem{B12-sensors-1607106}
Janani, A.S.; Grummett, T.S.; Lewis, T.W.; Fitzgibbon, S.P.; Whitham, E.M.; DelosAngeles, D.; Pope, K.J. Improved artefact removal from EEG using Canonical Correlation Analysis and spectral slope. \emph{J. Neurosci. Methods} \textbf{\boldmath{2018}}, \emph{298}, 1--15. [\href{http://doi.org/10.1016/j.jneumeth.2018.01.004}{CrossRef}]

\bibitem{B13-sensors-1607106}
Chen, X.; Peng, H.; Yu, F.; Wang, K. Independent vector analysis applied to remove muscle artifacts in EEG data. \emph{IEEE Trans. Instrum. Meas.} \textbf{\boldmath{2017}}, \emph{66}, 1770--1779. [\href{http://doi.org/10.1109/TIM.2016.2608479}{CrossRef}]

\bibitem{B14-sensors-1607106}
Chen, X.; Chen, Q.; Zhang, Y.; Wang, Z.J. A novel EEMD-CCA approach to removing muscle artifacts for pervasive EEG. \emph{IEEE Sens. J.} \textbf{\boldmath{2018}}, \emph{19}, 8420--8431. [\href{http://doi.org/10.1109/JSEN.2018.2872623}{CrossRef}]

\bibitem{B15-sensors-1607106}
Minguillon, J.; Lopez-Gordo, M.A.; Pelayo, F. Trends in EEG-BCI for daily-life: Requirements for artifact removal. \emph{Biomed. Signal Process. Control.} \textbf{\boldmath{2017}}, \emph{31}, 407--418. [\href{http://doi.org/10.1016/j.bspc.2016.09.005}{CrossRef}]

\bibitem{B16-sensors-1607106}
Fitzgibbon, S.P.; Lewis, T.W.; Powers, D.M.; Whitham, E.W.; Willoughby, J.O.; Pope, K.J. Surface laplacian of central scalp electrical signals is insensitive to muscle contamination. \emph{IEEE Trans. Biomed. Eng.} \textbf{\boldmath{2012}}, \emph{60}, 4--9. [\href{http://doi.org/10.1109/TBME.2012.2195662}{CrossRef}] [\href{http://www.ncbi.nlm.nih.gov/pubmed/22542648}{PubMed}]

\bibitem{B17-sensors-1607106}
Yong, X.; Ward, R.K.; Birch, G.E. Artifact removal in EEG using morphological component analysis. In Proceedings of the 2009 IEEE International Conference on Acoustics, Speech and Signal Processing, Taipei, Taiwan, 19--24 April 2009; pp. 345--348.

\bibitem{B18-sensors-1607106}
Bhardwaj, S.; Jadhav, P.; Adapa, B.; Acharyya, A.; Naik, G.R. Online and automated reliable system design to remove blink and muscle artefact in EEG. In Proceedings of the 2015 37th Annual International Conference of the IEEE Engineering in Medicine and Biology Society (EMBC), Milan, Italy, 25--29 August 2015; pp. 6784--6787.

\bibitem{B19-sensors-1607106}
Sadiq, M.T.; Siuly, S.; Ateeq Ur, R. Evaluation of power spectral and machine learning techniques for the development of \mbox{subject-specific} BCI. In \emph{Artificial Intelligence-Based Brain-Computer Interface}; Academic Press: Cambridge, MA, USA, \mbox{2022; pp. 99--120.}

\bibitem{B20-sensors-1607106}
Yu, X.; Aziz, M.Z.; Sadiq, M.T.; Jia, K.; Fan, Z.; Xiao, G. Computerized Multidomain EEG Classification System: A New Paradigm. \emph{IEEE J. Biomed. Health Inform.} \textbf{\boldmath{2022}}. [\href{http://doi.org/10.1109/JBHI.2022.3151570}{CrossRef}] [\href{http://www.ncbi.nlm.nih.gov/pubmed/35157605}{PubMed}]

\bibitem{B21-sensors-1607106}
Bakshi, B.R. Multiscale PCA with application to multivariate statistical process monitoring. \emph{AICHE J.} \textbf{\boldmath{1998}}, \emph{7}, 1596--1610. [\href{http://doi.org/10.1002/aic.690440712}{CrossRef}]

\bibitem{B22-sensors-1607106}
Mijović, B.; De Vos, M.; Gligorijević, I.; Taelman, J.; Van Huffel, S. Source separation from single-channel recordings by combining empirical-mode decomposition and independent component analysis. \emph{IEEE Trans. Biomed. Eng.} \textbf{\boldmath{2010}}, \emph{57}, 2188--2196. [\href{http://doi.org/10.1109/TBME.2010.2051440}{CrossRef}]

\bibitem{B23-sensors-1607106}
Sweeney, K.T.; McLoone, S.F.; Ward, T.E. The use of ensemble empirical mode decomposition with canonical correlation analysis as a novel artifact removal technique. \emph{IEEE Trans. Biomed. Eng.} \textbf{\boldmath{2012}}, \emph{60}, 97--105. [\href{http://doi.org/10.1109/TBME.2012.2225427}{CrossRef}]

\bibitem{B24-sensors-1607106}
Chen, X.; Liu, A.; Peng, H.; Ward, R.K. A preliminary study of muscular artifact cancellation in single-channel EEG. \emph{Sensors} \textbf{\boldmath{2014}}, \emph{14}, 18370--18389. [\href{http://doi.org/10.3390/s141018370}{CrossRef}]

\bibitem{B25-sensors-1607106}
Chen, X.; He, C.; Peng, H. Removal of muscle artifacts from single-channel EEG based on ensemble empirical mode decomposition and multiset canonical correlation analysis. \emph{J. Appl. Math.} \textbf{\boldmath{2014}}, \emph{2014}, 261347. [\href{http://doi.org/10.1155/2014/261347}{CrossRef}]

\bibitem{B26-sensors-1607106}
Zeng, K.; Chen, D.; Ouyang, G.; Wang, L.; Liu, X.; Li, X. An EEMD-ICA approach to enhancing artifact rejection for noisy multivariate neural data. \emph{IEEE Trans. Neural Syst. Rehabil. Eng.} \textbf{\boldmath{2015}}, \emph{24}, 630--638. [\href{http://doi.org/10.1109/TNSRE.2015.2496334}{CrossRef}] [\href{http://www.ncbi.nlm.nih.gov/pubmed/26552089}{PubMed}]

\bibitem{B27-sensors-1607106}
Chen, X.; Xu, X.; Liu, A.; McKeown, M.J.; Wang, Z.J. The use of multivariate EMD and CCA for denoising muscle artifacts from few-channel EEG recordings. \emph{IEEE Trans. Instrum. Meas.} \textbf{\boldmath{2017}}, \emph{67}, 359--370. [\href{http://doi.org/10.1109/TIM.2017.2759398}{CrossRef}]

\bibitem{B28-sensors-1607106}
Xu, X.; Chen, X.; Zhang, Y. Removal of muscle artefacts from few-channel EEG recordings based on multivariate empirical mode decomposition and independent vector analysis. \emph{Electron. Lett.} \textbf{\boldmath{2018}}, \emph{54}, 866--868. [\href{http://doi.org/10.1049/el.2018.0191}{CrossRef}]

\bibitem{B29-sensors-1607106}
Maddirala, A.K.; Shaik, R.A. Separation of sources from single-channel EEG signals using independent component analysis. \emph{IEEE Trans. Instrum. Meas.} \textbf{\boldmath{2017}}, \emph{67}, 382--393. [\href{http://doi.org/10.1109/TIM.2017.2775358}{CrossRef}]

\bibitem{B30-sensors-1607106}
Dora, C.; Patro, R.N.; Rout, S.K.; Biswal, P.K.; Biswal, B. Adaptive SSA Based Muscle Artifact Removal from Single Channel EEG using Neural Network Regressor. \emph{IRBM} \textbf{\boldmath{2021}}, \emph{42}, 324--333. [\href{http://doi.org/10.1016/j.irbm.2020.08.002}{CrossRef}]

\bibitem{B31-sensors-1607106}
Liu, Y.; Zhou, Y.; Lang, X.; Liu, Y.; Zheng, Q.; Zhang, Y.; Jiang, X.; Zhang, L.; Tang, J.; Dai, Y. An Efficient and Robust Muscle Artifact Removal Method for Few-Channel EEG. \emph{IEEE Access} \textbf{\boldmath{2019}}, \emph{7}, 176036--176050. [\href{http://doi.org/10.1109/ACCESS.2019.2957401}{CrossRef}]

\bibitem{B32-sensors-1607106}
Li, Y.; Wang, P.T.; Vaidya, M.P.; Liu, C.Y.; Slutzky, M.W.; Do, A.H. Electromyogram (EMG) Removal by Adding Sources of EMG (ERASE)---A novel ICA-based algorithm for removing myoelectric artifacts from EEG. \emph{Front. Neurosci.} \textbf{\boldmath{2021}}, \emph{14}, 1408. [\href{http://doi.org/10.3389/fnins.2020.597941}{CrossRef}]

\bibitem{B33-sensors-1607106}
Chen, X.; Xu, X.; Liu, A.; Lee, S.; Chen, X.; Zhang, X.; McKeown, M.J.; Wang, Z.J. Removal of muscle artifacts from the EEG: \mbox{A review} and recommendations. \emph{IEEE Sens. J.} \textbf{\boldmath{2019}}, \emph{19}, 5353--5368. [\href{http://doi.org/10.1109/JSEN.2019.2906572}{CrossRef}]

\bibitem{B34-sensors-1607106}
Phadikar, S.; Sinha, N.; Ghosh, R. Automatic eye blink artifact removal from eeg signal using wavelet transform with heuristically optimized threshold. \emph{IEEE J. Biomed. Health Inform.} \textbf{\boldmath{2021}}, \emph{25}, 475--484. [\href{http://doi.org/10.1109/JBHI.2020.2995235}{CrossRef}]

\bibitem{B35-sensors-1607106}
Bajaj, N.; Carri{\fontencoding{T5}\selectfont{\'o}}n, J.R.; Bellotti, F.; Berta, R.; De Gloria, A. Automatic and tunable algorithm for EEG artifact removal using wavelet decomposition with applications in predictive modeling during auditory tasks. \emph{Biomed. Signal Process. Control.} \textbf{\boldmath{2020}}, \mbox{\emph{55}, 101624. [\href{http://doi.org/10.1016/j.bspc.2019.101624}{CrossRef}]}

\bibitem{B36-sensors-1607106}
Zhang, H.; Wei, C.; Zhao, M.; Wu, H.; Liu, Q. A novel convolutional neural network model to remove muscle artifacts from EEG. In Proceedings of the IEEE International Conference on Acoustics, Speech and Signal Processing (ICASSP), Virtual Conference, 6--12 June 2021; pp. 1265--1269.

\bibitem{B37-sensors-1607106}
Buades, A.; Coll, B.; Morel, J.M. A review of image denoising algorithms, with a new one. \emph{Multiscale Model. Simul.} \textbf{\boldmath{2005}}, \mbox{\emph{4}, 490--530. [\href{http://doi.org/10.1137/040616024}{CrossRef}]}

\bibitem{B38-sensors-1607106}
Eltrass, A.S.; Ghanem, N.H. A new automated multi-stage system of non-local means and multi-kernel adaptive filtering techniques for EEG noise and artifacts suppression. \emph{J. Neural Eng.} \textbf{\boldmath{2021}}, \emph{18}, 036023. [\href{http://doi.org/10.1088/1741-2552/abe397}{CrossRef}] [\href{http://www.ncbi.nlm.nih.gov/pubmed/33545699}{PubMed}]

\bibitem{B39-sensors-1607106}
Ghosh, R.; Sinha, N.; Biswas, S.K. Automated eye blink artefact removal from EEG using support vector machine and autoencoder. \emph{IET Signal Process.} \textbf{\boldmath{2019}}, \emph{13}, 141--148. [\href{http://doi.org/10.1049/iet-spr.2018.5111}{CrossRef}]

\bibitem{B40-sensors-1607106}
Cortes, C.; Vapnik, V. Support-vector networks. \emph{Mach. Learn.} \textbf{\boldmath{1995}}, \emph{20}, 273--297. [\href{http://doi.org/10.1007/BF00994018}{CrossRef}]

\bibitem{B41-sensors-1607106}
\textls[-20]{Pisner, D.A.; Schnyer, D.M. Support vector machine. In \emph{Machine Learning}; Academic Press: Cambridge, MA, USA, \mbox{2020; pp. 101--121.}}

\bibitem{B42-sensors-1607106}
Ghaderpour, E.; Pagiatakis, S.D.; Hassan, Q.K. A survey on change detection and time series analysis with applications. \emph{Appl. Sci.} \textbf{\boldmath{2021}}, \emph{11}, 6141. [\href{http://doi.org/10.3390/app11136141}{CrossRef}]

\bibitem{B43-sensors-1607106}
Ting, W.; Guo-Zheng, Y.; Bang-Hua, Y.; Hong, S. EEG feature extraction based on wavelet packet decomposition for brain computer interface. \emph{Measurement} \textbf{\boldmath{2008}}, \emph{41}, 618--625. [\href{http://doi.org/10.1016/j.measurement.2007.07.007}{CrossRef}]

\bibitem{B44-sensors-1607106}
Jaffery, Z.A.; Ahmad, K.; Sharma, P. Selection of optimal decomposition level based on entropy for speech denoising using wavelet packet. \emph{J. Bioinform. Intell. Control.} \textbf{\boldmath{2012}}, \emph{1}, 196--202. [\href{http://doi.org/10.1166/jbic.2013.1026}{CrossRef}]

\bibitem{B45-sensors-1607106}
Mirjalili, S.; Mirjalili, S.M.; Lewis, A. Grey wolf optimizer. \emph{Adv. Eng. Softw.} \textbf{\boldmath{2014}}, \emph{69}, 46--61. [\href{http://doi.org/10.1016/j.advengsoft.2013.12.007}{CrossRef}]

\bibitem{B46-sensors-1607106}
Ghosh, R.; Sinha, N.; Biswas, S.K.; Phadikar, S. A modified grey wolf optimization based feature selection method from EEG for silent speech classification. \emph{J. Inf. Optim. Sci.} \textbf{\boldmath{2019}}, \emph{40}, 1639--1652. [\href{http://doi.org/10.1080/02522667.2019.1703262}{CrossRef}]

\bibitem{B47-sensors-1607106}
Van De Ville, D.; Kocher, M. SURE-based non-local means. \emph{IEEE Signal Process. Lett.} \textbf{\boldmath{2009}}, \emph{16}, 973--976. [\href{http://doi.org/10.1109/LSP.2009.2027669}{CrossRef}]

\bibitem{B48-sensors-1607106}
Singh, P.; Pradhan, G.; Shahnawazuddin, S. Denoising of ECG signal by non-local estimation of approximation coefficients in DWT. \emph{Biocybern. Biomed. Eng.} \textbf{\boldmath{2017}}, \emph{37}, 599--610. [\href{http://doi.org/10.1016/j.bbe.2017.06.001}{CrossRef}]

\bibitem{B49-sensors-1607106}
Ghanem, N.H.; Eltrass, A.S.; Ismail, N.H. Investigation of EEG noise and artifact removal by patch-based and kernel adaptive filtering techniques. In Proceedings of the 2018 IEEE International Symposium on Medical Measurements and Applications (MeMeA), Rome, Italy, 11--13 June 2018; pp. 1--5.

\bibitem{B50-sensors-1607106}
Phadikar, S.; Sinha, N.; Ghosh, R. Automatic EEG eyeblink artefact identification and removal technique using independent component analysis in combination with support vector machines and denoising autoencoder. \emph{IET Signal Process.} \textbf{\boldmath{2020}}, \mbox{\emph{14}, 396--405. [\href{http://doi.org/10.1049/iet-spr.2020.0025}{CrossRef}]}

\bibitem{B51-sensors-1607106}
Ghosh, R.; Deb, N.; Sengupta, K.; Phukan, A.; Choudhury, N.; Kashyap, S.; Dutta, P. SAM 40: Dataset of 40 subject EEG recordings to monitor the induced-stress while performing Stroop color-word test, arithmetic task, and mirror image recognition task. \emph{Data Brief} \textbf{\boldmath{2022}}, \emph{40}, 107772. [\href{http://doi.org/10.1016/j.dib.2021.107772}{CrossRef}] [\href{http://www.ncbi.nlm.nih.gov/pubmed/35036481}{PubMed}]

\bibitem{B52-sensors-1607106}
Chen, X.; Liu, A.; Chiang, J.; Wang, Z.J.; McKeown, M.J.; Ward, R.K. Removing muscle artifacts from EEG data: Multichannel or single-channel techniques? \emph{IEEE Sens. J.} \textbf{\boldmath{2016}}, \emph{16}, 1986--1997. [\href{http://doi.org/10.1109/JSEN.2015.2506982}{CrossRef}]

\end{thebibliography}
\end{document}